\documentclass[twocolumn]{autart}    % Enable this line and disable the
\usepackage{graphicx}          % Include this line if your
\usepackage{amsmath}
\usepackage{amssymb}
\usepackage{color}
\usepackage{hyperref}

\def \be {\begin{equation}}

\def \ee {\end{equation}}

\def \nn {\nonumber}

\begin{document}

\begin{frontmatter}

\title{Construction of Differential-Cascaded Structures for Control of Robot Manipulators} % Title, preferably not more

\thanks[footnoteinfo]{Corresponding author
Hanlei Wang.
}

\author{Hanlei Wang}\ead{hlwang.bice@gmail.com}

\address{Science and Technology on Space Intelligent Control Laboratory, Beijing Institute of
Control Engineering, Beijing 100094, China}

                             % keyword list or with the
                                          % help of the Automatica
                                          % keyword wizard

\begin{abstract}                          % Abstract of not more than 200 words.
This paper focuses on the construction of differential-cascaded structures for control of nonlinear robot manipulators subjected to disturbances and unavailability of partial information of the desired trajectory. The proposed differential-cascaded structures rely on infinite differential series to handle the robustness with respect to time-varying disturbances and the partial knowledge of the desired trajectories for nonlinear robot manipulators. The long-standing problem of reliable adaptation in the presence of sustaining disturbances is solved by the proposed forwardstepping control with forwardstepping adaptation, and stacked reference dynamics yielding adaptive differential-cascaded structures have been proposed to facilitate the forwardstepping adaptation to both the uncertainty of robot dynamics and that of the frequencies of disturbances. A distinctive point of the proposed differential-cascaded approach is that the reference dynamics for design and analysis involve high-order quantities, but via degree-reduction implementation of the reference dynamics, the control typically involves only the low-order quantities, thus facilitating its applications to control of most physical systems. Our result relies on neither the explicit estimation of the disturbances or derivative and second derivative of the desired position nor the solutions to linear/nonlinear regulator equations, and the employed essential element is a differential-cascaded structure governing robot dynamics.
\end{abstract}

\end{frontmatter}

\section{Introduction}

The control of nonlinear robot manipulators has a long and versatile tradition, and numerous controllers have been proposed in various contexts (see, e.g., \cite{Takegaki1981_ASME,Craig1987_IJRR,Slotine1987_IJRR,Middleton1988_SCL,Slotine1989_AUT,Spong1989_SCL,Ortega1989_AUT,Berghuis1993_SCL,Cheah2006_IJRR,Liu2006_TRO,Wang2015_AUT,Wang2017_TAC,Wang2020_TAC}). In parallel, we witness the formulation of design and analysis approaches for general nonlinear systems such as backstepping \cite{Krstic1995_Book,Kokotovic2001_AUT}, forwarding \cite{Teel1992_IFAC,Sepulchre1997_AUT}, and the immersion and invariance (I\&I) approach \cite{Astolfi2003_TAC}. In particular, the control developed for robot manipulators exhibits implicitly or explicitly the nature of a design methodology like backstepping, and this is particularly reflected in the design of reference velocity and that of reference acceleration via a differential operation (see, e.g., \cite{Slotine1987_IJRR,Cheah2006_IJRR,Wang2017_TAC}). This procedure of specifying the reference velocity and acceleration can also be classified as the conventional order-reduction approach, which generally yields closed-loop dynamics that are cascaded; see \cite{Astolfi2003_TAC} for some further details. Recently, a differential-cascaded approach (also referred to as forwardstepping) \cite{Wang2019_ACC,Wang2020_arXiv} has been formulated; this approach has been demonstrated to be instrumental as handling time-varying delay (and switching topology) for networked Lagrangian systems or teleoperators \cite{Wang2020_AUT,Wang2020b_AUT,Wang2019_CDC}. Differing from the traditional order reduction paradigm \cite{Krstic1995_Book,Astolfi2003_TAC}, this differential-cascaded approach is an order invariance or increment one, and the resultant closed-loop dynamics are typically differential-cascaded. Due to the design freedom permitted by the differential-cascaded paradigm, the connection between controlled robot dynamics with uncertainty and linear dynamics, which is a long-standing intractable problem, has been established in \cite{Wang2021_arXiv}.

Trajectory tracking and robustness with respect to disturbances are two important topics of sustaining interest in control systems. In the standard proportional-integral-derivative (PID) control, the two problems are inherently accommodated in a unified way, and PID control, due to its particular structure, enjoys both the adaptability and robustness. The limitation of PID control is also well recognized, namely the integral action can only handle a constant or slowly time-varying disturbance and the PID control can generally handle the regulation problem or tracking of a slowly time-varying trajectory. To alleviate this limitation, many approaches have been presented with a versatility of features in design and analysis where the two problems are typically accommodated in either very different ways or in a unified way such as using the standard internal model approach \cite{Isidori1990_TAC,Byrnes2000_IJRNC}. The current internal model approach and many other approaches based on the estimation of the disturbances are, however, not very aligned with the essence of the standard PID control due to either the constraint of solving a linear/nonlinear regulator equation or that of the explicit estimation of disturbances; the structural conciseness and compactness of PID control also distinguishes it from most available approaches. In this sense, the standard sliding control exploiting discontinuous signum functions may be considered to be aligned with PID control except that the sliding control is subjected to the well-known chattering problem (see, e.g., \cite{Slotine1991_Book}); some attempts to alleviate this issue occur in, e.g., \cite{Xian2004_TAC,Patre2010_AUT} at the expense of semi-global stability and complexity of the developed control. On the other hand, in the distributed trajectory tracking problem, the signum-function-based action has been exploited in the context of linear systems with undirected topology (see, e.g., \cite{Cao2012_TAC}) or nonlinear Lagrangian systems with directed topology (see, e.g., \cite{Wang2020_arXiv}) where the explicit estimation of the trajectory of the leader is not required. An important issue in trajectory tracking for robot manipulators (and also general nonlinear systems) is that the derivative and twice derivative of the trajectory are generally required to realize the asymptotic tracking. While the availability of the derivative and twice derivative of the trajectory may not be so challenging in the traditional factory automation with a trajectory planner, this becomes intractable in the case of tracking a moving target in which case only the measurement of the desired position is typically possible. The unavailability of such information renders the asymptotic tracking to be impossible in the case of a fast time-varying desired trajectory. The ad hoc solution given in \cite[p.~195]{Slotine1991_Book} in the context without involving the derivative and second derivative of the desired trajectory, however, cannot guarantee the asymptotic tracking.

In this paper, we investigate the design of differential-cascaded structures for control of nonlinear robot manipulators, and the tracking of a time-varying trajectory and robustness with respect to unknown time-varying disturbances are accommodated. The main element for solving the two problems in a unified way is a differential-cascaded design using high-order reference dynamics or forwardstepping design \cite{Wang2019_ACC,Wang2020_arXiv}, which yields differential-cascaded structures. The presented approach has some similarity to the internal model approach in the sense of not relying on the explicit estimation of the disturbances or the derivative and twice derivative of the desired trajectory. The distinctive point of our methodology lies in the construction of differential-cascaded structures to yield solutions in the sense of the limit of the order or approximation with a remainder that can be  systematically decreased (due to the use of differential series), in contrast with the internal model approach that relies on solutions to linear or nonlinear regulator equations (see, e.g., \cite{Isidori1990_TAC,Byrnes2000_IJRNC,Jayawardhana2008_AUT,Wu2020_CCC}) and also with the other approaches based upon disturbance estimation.
Our result provides solutions to the long-standing problem of reliable adaptation to uncertainty of the system dynamics in the presence of sustaining disturbances via the example of nonlinear robot manipulators, and the proposed adaptation is referred to as forwardstepping adaptation, differing from most adaptation in the literature (e.g., \cite{Slotine1991_Book,Ioannou1996_Book,Slotine1987_IJRR,Wu2019_ICCA,Lu2019_AUT}) that tends to be unreliable or even fragile as encountering disturbances. The weak robustness or reliability of the conventional adaptive control is recognized to be a fundamental limitation in practical applications. The presented forwardstepping adaptation, via exploiting the differential-cascaded structure yielded by the proposed framework, ensures the reliability of the system in the presence of disturbances.  We also present a connection between the differential-cascaded framework and the standard PID control, and via a suitable reformulation, it is demonstrated that the reformulation of the PID control yields a differential-cascaded structure with order one.

\section{Equations of Motion of Robot Manipulators}

The equations of motion of a robot manipulator can be written as \cite{Slotine1991_Book,Spong2006_Book}
\be
\label{eq:1}
M(q)\ddot q+C(q,\dot q)\dot q+g(q)=\tau+\tau^\ast
\ee
where $q\in R^n$ is the joint position, $M(q)\in R^{n\times n}$ is the inertia matrix, $C(q,\dot q)\in R^{n\times n}$ is the Coriolis and centrifugal matrix, $g(q)\in R^n$ is the gravitational torque, $\tau\in R^n$ is the joint control torque, and $\tau^\ast\in R^n$ is the reference torque input (e.g., a disturbance). Three standard properties associated with the dynamics (\ref{eq:1}) are formulated as follows (see, e.g., \cite{Slotine1991_Book,Spong2006_Book}).

\emph{Property 1:} The inertia matrix $M(q)$ is symmetric and uniformly positive definite.

\emph{Property 2:} The Coriolis and centrifugal matrix $C(q,\dot q)$ can be suitably chosen such that $\dot M(q)-2C(q,\dot q)$ is skew-symmetric.

\emph{Property 3:} The dynamics (\ref{eq:1}) depend linearly on a constant parameter vector $\vartheta$, yielding the result that $M(q)\dot \zeta+C(q,\dot q)\zeta+g(q)=Y(q,\dot q,\zeta,\dot\zeta)\vartheta$ where $\zeta\in R^n$ is a differentiable vector and $Y(q,\dot q,\zeta,\dot \zeta)$ is a regressor matrix.

\section{Forwardstepping Control}

\subsection{Design and Analysis}

Let $q_d\in R^n$ denote the desired position, and assume that  $q_d$ is arbitrary times differentiable and that $q_d$, $\dot q_d$, and $\ddot q_d$ are bounded. We consider the trajectory tracking problem of robot manipulators without the knowledge of $\dot q_d$ and $\ddot q_d$ and without disturbances (namely $\tau^\ast=0$). For this purpose, we introduce the following reference dynamics with order $\ell+1$
\be
\label{eq:2}
\frac{d^\ell z}{dt^\ell}=-\alpha_\ell \frac{d^\ell q}{dt^\ell}-\alpha_{\ell-1} \frac{d^{\ell-1} \Delta q}{dt^{\ell-1}}-\dots-\alpha_0\Delta q
\ee
where $\Delta q=q-q_d$ is the joint position tracking error, $\alpha_0,\dots,\alpha_\ell$ are chosen such that $w^{\ell+1}+\alpha_\ell w^\ell+\dots+\alpha_0$ with $w$ being the free variable is a Hurwitz polynomial, and $\ell\ge 1$. The vector $z$ generated by (\ref{eq:2}) acts as the reference velocity. Define
\be
\label{eq:3}
s=\dot q-z.
\ee
The torque input is specified by the following adaptive control
\begin{align}
\label{eq:4}
\tau=&-K s+Y(q,\dot q,z,\dot z)\hat\vartheta\\
\label{eq:5}
\dot{\hat\vartheta}=&-\Gamma Y^T(q,\dot q,z,\dot z)s
\end{align}
where $K$ and $\Gamma$ are symmetric positive definite matrices, and $\hat\vartheta$ is the estimate of $\vartheta$. The adaptive controller given as (\ref{eq:4}) and (\ref{eq:5}) has the same structure as the standard one in \cite{Slotine1987_IJRR}. We can also apply the other adaptive approaches (e.g., \cite{Wang2021_arXiv}) to specify the control torque. Due to the introduction of the $\ell+1$-th-order reference dynamics (\ref{eq:2}), we refer to the presented adaptive control as forwardstepping adaptive control.

The closed-loop dynamics can be described by a differential-cascaded system with degree $\ell$ as
\be
\label{eq:6}
\begin{cases}
\frac{d^{\ell+1} q}{dt^{\ell+1}}=-\alpha_\ell \frac{d^\ell q}{dt^\ell}-\alpha_{\ell-1} \frac{d^{\ell-1} \Delta q}{dt^{\ell-1}}-\dots-\alpha_0\Delta q+\frac{d^\ell s}{dt^\ell}\\
M(q)\dot s+C(q,\dot q)s=-K s+Y(q,\dot q,z,\dot z)\Delta\vartheta\\
\dot{\hat\vartheta}=-\Gamma Y^T(q,\dot q,z,\dot z)s
\end{cases}
\ee
where $\Delta\vartheta=\hat\vartheta-\vartheta$.

\emph{Theorem 1:} Suppose that the desired position $q_d$ satisfies the property that $\frac{d^{\ell}q_d}{dt^{\ell}}=0$ and $\frac{d^{\ell+1}q_d}{dt^{\ell+1}}=0$. Then, the forwardstepping adaptive controller given as (\ref{eq:2}), (\ref{eq:4}), and (\ref{eq:5}) ensures the convergence of the joint tracking errors, i.e., $\Delta q\to 0$ and $\Delta \dot q\to0 $ as $t\to\infty$.

\emph{Proof:} Following the typical practice (see, e.g., \cite{Slotine1987_IJRR,Ortega1989_AUT}), we consider the Lyapunov-like function candidate $V=(1/2)s^TM(q)s+(1/2)\Delta\vartheta^T \Gamma^{-1}\Delta \vartheta$, and the derivative of $V$ along the trajectories of the system can be written as (using Property 2) $\dot V=-s^T Ks\le 0$. This implies that $s\in\mathcal L_2\cap\mathcal L_\infty$ and $\hat\vartheta\in\mathcal L_\infty$. With $\frac{d^{\ell}q_d}{dt^{\ell}}=0$ and $\frac{d^{\ell+1}q_d}{dt^{\ell+1}}=0$, the first subsystem of (\ref{eq:6}) can be rewritten as
\be
\label{eq:7}
\frac{d^{\ell+1} \Delta q}{dt^{\ell+1}}=-\alpha_\ell \frac{d^\ell \Delta q}{dt^\ell}-\dots-\alpha_0\Delta q+\frac{d^\ell s}{dt^\ell}.
\ee
From the standard input-output properties of linear systems (see, e.g., \cite[p.~59]{Desoer1975_Book}), we have from (\ref{eq:7}) that $\Delta q\in\mathcal L_2\cap\mathcal L_\infty$, $\Delta \dot q\in\mathcal L_2\cap\mathcal L_\infty$, and $\Delta q\to 0$ as $t\to\infty$. Hence, $\dot q\in\mathcal L_\infty$. Using the relation $s=\dot q-z$ yields the result that $z\in\mathcal L_\infty$. We can further write (\ref{eq:7}) as
\begin{align}
\label{eq:8}
\frac{d^{\ell} \Delta z}{dt^{\ell}}=&-\alpha_\ell \frac{d^{\ell-1} \Delta z}{dt^{\ell-1}}-\dots-\alpha_1\Delta z-\alpha_0\Delta q\nn\\
&-\alpha_\ell\frac{d^{\ell-1}s}{dt^{\ell-1}}-\dots-\alpha_1 s
\end{align}
where $\Delta z=z-\dot q_d$. The state-space representation for (\ref{eq:8}) can be written as
\begin{align}
\label{eq:9}
\begin{cases}
\Delta \dot q=\Delta z+s\\
\quad\vdots\\
\frac{d}{dt}({\frac{d^{\ell-2} \Delta z}{dt^{\ell-2}}})={\frac{d^{\ell-1} \Delta z}{dt^{\ell-1}}}\\
\frac{d}{dt}(\frac{d^{\ell-1} \Delta z}{dt^{\ell-1}})=-\alpha_\ell \frac{d^{\ell-1} \Delta z}{dt^{\ell-1}}-\dots-\alpha_1\Delta z-\alpha_0\Delta q\\
\qquad-\alpha_\ell\frac{d^{\ell-1}s}{dt^{\ell-1}}-\dots-\alpha_1 s
\end{cases}
\end{align}
with $x=[\Delta q^T,\Delta z^T,\dots,(d^{\ell-1}\Delta z/dt^{\ell-1})^T]^T$ as the state and with $u_1=[s^T,0,\dots,0,-\alpha_1 s^T]^T$, $u_k=-\alpha_kd^{k-1}s/dt^{k-1}$, $k=2,\dots,\ell$ as the inputs. The system (\ref{eq:9}) with $u_k=0$, $k=1,\dots,\ell$ is exponentially stable since $w^{\ell+1}+\alpha_\ell w^\ell+\dots+\alpha_0$ is a Hurwitz polynomial. From the input-output properties of exponentially stable and strictly proper linear systems \cite[p.~59]{Desoer1975_Book}, we have that the part of $x$ due to the input $u_1$ is bounded. Consider the mapping from $u_k$ to $\Delta \dot z$ and the relative degree of $\Delta \dot z$ with respect to $-\alpha_k s$ is $\ell-1-(k-1)=\ell-k\ge0$, and thus this mapping is proper according to the standard linear system theory, $k=2,\dots,\ell$. Using the standard input-output properties of linear systems (see \cite[p.~59]{Desoer1975_Book} and \cite[p.~82]{Ioannou1996_Book}) yields the result that the portion of $\Delta \dot z$ due to the input $u_k$ is bounded, $k=2,\dots,\ell$. From the standard superposition principle for linear systems, we have that $\Delta \dot z\in\mathcal L_\infty$, and hence that $\dot z\in\mathcal L_\infty$. Based on the second subsystem of (\ref{eq:6}) and using Property 1, we have that $\dot s\in\mathcal L_\infty$, and hence $\ddot q\in\mathcal L_\infty$. This implies that $\Delta\ddot q\in\mathcal L_\infty$, and thus $\Delta \dot q$ is uniformly continuous. Using Barbalat's lemma \cite{Slotine1991_Book}, we have that $\Delta \dot q\to0$ as $t\to\infty$.
\hfill\text{\small $\blacksquare$}

\emph{Remark 1:} With the assumption that $d^\kappa q_d/dt^\kappa$, $\kappa=3,\dots,\ell-1$ are bounded, we can show via a similar procedure as in the proof of Theorem 1 that $d^{\kappa-1} z/dt^{\kappa-1}$ and $d^\kappa \Delta q/dt^\kappa$ $\kappa=3,\dots,\ell+1$ are bounded. However, this assumption is not required for ensuring the stability and convergence of the presented forwardstepping adaptive controller. In the case that $d^{\ell}q_d/dt^{\ell}$ and $d^{\ell+1}q_d/dt^{\ell+1}$ are nonzero and bounded, the convergence of the presented forwardstepping control cannot be ensured, yet the tracking errors $\Delta q$ and $\Delta \dot q$ can still be ensured to be bounded, which can be demonstrated via similar procedures as in the proof of Theorem 1.

\subsection{Degree-Reduction Implementation}

The reference dynamics (\ref{eq:2}) involve high-order quantities that are difficult or impossible to be measured, and following the practice in \cite{Wang2021_arXiv}, we use the degree reduction to obtain an implementation scheme for generating $z$ and $\dot z$ that only relies on the quantities $q$, $\dot q$, and $q_d$ (the derivative and twice derivative of $q_d$ are not involved). Specifically, we write a state-space representation of (\ref{eq:2}) as
\be
\begin{cases}
\dot z=\dot z\\
\quad\vdots\\
\frac{d^{\ell-1}z}{dt^{\ell-1}}=\frac{d^{\ell-1}z}{dt^{\ell-1}}\\
\frac{d^\ell z}{dt^\ell}=-\alpha_\ell \frac{d^\ell  q}{dt^\ell}-\dots-\alpha_0\Delta q.
\end{cases}
\ee
We apply the degree reduction concerning $-\alpha_\ell {d^\ell q}/{dt^\ell}$ and have that
\be
\begin{cases}
\dot z=\dot z\\
\quad\vdots\\
\frac{d^{\ell-2}z}{dt^{\ell-2}}=\frac{d^{\ell-2}z}{dt^{\ell-2}}\\
\frac{d^{\ell-1}z}{dt^{\ell-1}}=(\frac{d^{\ell-1}z}{dt^{\ell-1}}+\alpha_\ell\frac{d^{\ell-1}  q}{dt^{\ell-1}})-\alpha_\ell\frac{d^{\ell-1} q}{dt^{\ell-1}}\\
\frac{d}{dt}(\frac{d^{\ell-1}z}{dt^{\ell-1}}+\alpha_\ell\frac{d^{\ell-1}  q}{dt^{\ell-1}})=-\alpha_{\ell-1} \frac{d^{\ell-1} \Delta q}{dt^{\ell-1}}-\dots-\alpha_0\Delta q.
\end{cases}
\ee
Then, applying the degree reduction concerning the terms $-\alpha_\ell{d^{\ell-1} q}/{dt^{\ell-1}}$ and $-\alpha_{\ell-1}{d^{\ell-1} \Delta q}/{dt^{\ell-1}}$ yields
\be
\begin{cases}
\dot z=\dot z\\
\quad\vdots\\
\frac{d^{\ell-2}z}{dt^{\ell-2}}=(\frac{d^{\ell-2}z}{dt^{\ell-2}}+\alpha_{\ell} \frac{d^{\ell-2}  q}{dt^{\ell-2}})-\alpha_{\ell} \frac{d^{\ell-2}  q}{dt^{\ell-2}}\\
\frac{d}{dt}(\frac{d^{\ell-2}z}{dt^{\ell-2}}+\alpha_\ell\frac{d^{\ell-2}  q}{dt^{\ell-2}})\\=(\frac{d^{\ell-1}z}{dt^{\ell-1}}+\alpha_\ell\frac{d^{\ell-1}  q}{dt^{\ell-1}}+\alpha_{\ell-1} \frac{d^{\ell-2} \Delta q}{dt^{\ell-2}})\\
\quad-\alpha_{\ell-1} \frac{d^{\ell-2} \Delta q}{dt^{\ell-2}}\\
\frac{d}{dt}(\frac{d^{\ell-1}z}{dt^{\ell-1}}+\alpha_\ell\frac{d^{\ell-1}  q}{dt^{\ell-1}}+\alpha_{\ell-1} \frac{d^{\ell-2} \Delta q}{dt^{\ell-2}})\\
=-\alpha_{\ell-2} \frac{d^{\ell-2} \Delta q}{dt^{\ell-2}}-\dots-\alpha_0\Delta q.
\end{cases}
\ee
With this procedure being continued, we finally have a degree-reduction scheme given as
\be
\begin{cases}
\frac{d}{dt}z=(\dot z+\alpha_{\ell} \dot q-\alpha_{\ell-1}q_d)\\
\quad-\alpha_{\ell}\dot q+\alpha_{\ell-1}q_d\\
\frac{d}{dt}(\dot z+\alpha_\ell \dot q-\alpha_{\ell-1}q_d)\\=(\ddot z+\alpha_\ell\ddot q+\alpha_{\ell-1}\Delta \dot q-\alpha_{\ell-2}q_d)\\
\quad -\alpha_{\ell-1}\dot q+\alpha_{\ell-2}q_d\\
\quad\vdots\\
\frac{d}{dt}(\frac{d^{\ell-2}z}{dt^{\ell-2}}+\alpha_\ell\frac{d^{\ell-2}  q}{dt^{\ell-2}}+\dots+\alpha_3\Delta \dot q-\alpha_2 q_d)\\
=(\frac{d^{\ell-1}z}{dt^{\ell-1}}+\alpha_\ell\frac{d^{\ell-1} q}{dt^{\ell-1}}+\alpha_{\ell-1} \frac{d^{\ell-2} \Delta q}{dt^{\ell-2}}+\cdots\\
\quad+\alpha_2\Delta \dot q-\alpha_1 q_d)-\alpha_2\dot q+\alpha_1q_d\\
\frac{d}{dt}(\frac{d^{\ell-1}z}{dt^{\ell-1}}+\alpha_\ell\frac{d^{\ell-1}  q}{dt^{\ell-1}}+\alpha_{\ell-1} \frac{d^{\ell-2} \Delta q}{dt^{\ell-2}}+\cdots\\
\quad+\alpha_2\Delta \dot q-\alpha_1 q_d)
=-\alpha_1\dot q-\alpha_0\Delta q.
\end{cases}
\ee
The implementation of the reference dynamics (\ref{eq:2}) can thus be conducted only using the quantities $q$, $\dot q$, and $q_d$.

\emph{Remark 2:} In the context that the desired velocity $\dot q_d$ is additionally available, the reference dynamics (\ref{eq:2}) can be redefined as
\be
\label{eq:14}
\frac{d^\ell z}{dt^\ell}=-\alpha_\ell \frac{d^\ell \Delta q}{dt^\ell}-\alpha_{\ell-1} \frac{d^{\ell-1} \Delta q}{dt^{\ell-1}}-\dots-\alpha_0\Delta q
\ee
and a degree-reduction implementation scheme for (\ref{eq:14}) can be written as
\be
\begin{cases}
\frac{d}{dt}z=(\dot z+\alpha_{\ell}\Delta \dot q)-\alpha_{\ell}\Delta\dot q\\
\frac{d}{dt}(\dot z+\alpha_\ell \Delta\dot q)=(\ddot z+\alpha_\ell\Delta\ddot q+\alpha_{\ell-1}\Delta \dot q) -\alpha_{\ell-1}\Delta\dot q\\
\quad\vdots\\
\frac{d}{dt}(\frac{d^{\ell-2}z}{dt^{\ell-2}}+\alpha_\ell\frac{d^{\ell-2} \Delta q}{dt^{\ell-2}}+\dots+\alpha_3\Delta \dot q)\\
=(\frac{d^{\ell-1}z}{dt^{\ell-1}}+\alpha_\ell\frac{d^{\ell-1}\Delta q}{dt^{\ell-1}}+\alpha_{\ell-1} \frac{d^{\ell-2} \Delta q}{dt^{\ell-2}}+\dots+\alpha_2\Delta \dot q)\\
\quad-\alpha_2\Delta\dot q\\
\frac{d}{dt}(\frac{d^{\ell-1}z}{dt^{\ell-1}}+\alpha_\ell\frac{d^{\ell-1}  \Delta q}{dt^{\ell-1}}+\alpha_{\ell-1} \frac{d^{\ell-2} \Delta q}{dt^{\ell-2}}+\dots+\alpha_2\Delta \dot q)\\
=-\alpha_1\Delta\dot q-\alpha_0\Delta q.
\end{cases}
\ee
For the common case that $q_d$, $\dot q_d$, and $\ddot q_d$ are all available, the reference dynamics can be defined as
\be
\label{eq:16}
\frac{d^\ell z}{dt^\ell}=\frac{d^{\ell+1} q_d}{dt^{\ell+1}}-\alpha_\ell \frac{d^\ell \Delta q}{dt^\ell}-\alpha_{\ell-1} \frac{d^{\ell-1} \Delta q}{dt^{\ell-1}}-\dots-\alpha_0\Delta q
\ee
with a degree-reduction implementation scheme being given as
\be
\begin{cases}
\frac{d}{dt}z=(\dot z-\ddot q_d+\alpha_{\ell}\Delta \dot q)+\ddot q_d-\alpha_{\ell}\Delta\dot q\\
\frac{d}{dt}(\dot z-\ddot q_d+\alpha_\ell \Delta\dot q)\\
=(\ddot z-\dddot q_d+\alpha_\ell\Delta\ddot q+\alpha_{\ell-1}\Delta \dot q) -\alpha_{\ell-1}\Delta\dot q\\
\quad\vdots\\
\frac{d}{dt}(\frac{d^{\ell-2}z}{dt^{\ell-2}}-\frac{d^{\ell-1} q_d}{dt^{\ell-1}}+\alpha_\ell\frac{d^{\ell-2}  \Delta q}{dt^{\ell-2}}+\dots+\alpha_3\Delta \dot q)\\
=(\frac{d^{\ell-1}z}{dt^{\ell-1}}-\frac{d^{\ell} q_d}{dt^{\ell}}+\alpha_\ell\frac{d^{\ell-1}\Delta q}{dt^{\ell-1}}+\dots+\alpha_2\Delta \dot q)-\alpha_2\Delta\dot q\\
\frac{d}{dt}(\frac{d^{\ell-1}z}{dt^{\ell-1}}-\frac{d^{\ell} q_d}{dt^{\ell}}+\alpha_\ell\frac{d^{\ell-1}  \Delta q}{dt^{\ell-1}}+\dots+\alpha_2\Delta \dot q)\\
=-\alpha_1\Delta\dot q-\alpha_0\Delta q.
\end{cases}
\ee
This degree-reduction scheme extends the specific examples in \cite{Wang2021_arXiv} to the general case.

\section{Forwardstepping Control Using Infinite Differential Series}

In a general context with the desired trajectory containing periodical information, the assumption in Theorem 1 can only hold as $\ell\to\infty$. The disturbance torque $\tau^\ast$ has similar properties as the desired trajectory, and the presence of the disturbance renders the use of the typical adaptation as in the previous section hard to be justified due to the well-recognized robustness issue (see, e.g., \cite{Slotine1991_Book}). For handling these problems in a unified way, we introduce what is referred to as infinite differential series.

\subsection{Infinite Differential Series}

In the context of forwardstepping control, the (infinite) differential series (in the finite case as in the previous section, we may refer to it as the sum of a differential sequence) becomes an important element for solving many important control problems. Consider a function $f(t)$ that is arbitrary times differentiable and a sequence of numbers $\alpha_0,\dots,\alpha_{m^\ast}$. Then, $\alpha_0 f(t),\alpha_1df(t)/dt,\dots,\alpha_{m^\ast}d^{m^\ast}f(t)/dt^{m^\ast}$ is referred to as a differential sequence (with respect to the function $f(t)$); as $m^\ast\to\infty$, this sequence is referred to as an infinite differential sequence; $\lim_{m^\ast\to\infty}\Sigma_{k=0}^{m^\ast}\alpha_kd^k f(t)/dt^k$ is referred to as an infinite differential series or differential series. Differential sequences and differential series can be considered as a particular class of general sequences and series as involving the derivative and high-order derivatives of a function.

In the traditional calculus, the object of interest is a function; in control systems, we focus on dynamical systems or differential equations. In this latter context, we are interested in differential series in defining the reference dynamics. For instance, consider an equation of an infinite differential series given as
\be
\sum_{k=0}^\infty \alpha_k\frac{d^k\Delta q}{dt^k}=\lim_{\ell\to\infty}\sum_{k=0}^{\ell} \alpha_k\frac{d^k\Delta q}{dt^k}=0,
\ee
and we view it as an equation concerning a series of a function, its derivative, and its high-order derivatives. Hence, it can also be referred to as an infinite-order differential equation while the left side is referred to as an infinite differential series defined above.

In the linear time-invariant case, exponentially stable polynomials or Hurwitz polynomials play an important role, which yield the following differential equation
\be
\sum_{k=0}^{\ell} \alpha_k\frac{d^k\Delta q}{dt^k}=0
\ee
with $\alpha_0,\dots,\alpha_\ell$ being chosen such that $\alpha_\ell w^\ell+\dots+\alpha_0$ is a Hurwitz polynomial. As $\ell$ approaches infinity, we obtain an equation of an infinite differential series (or concisely a differential series) that is exponentially stable.

\subsection{Forwardstepping Control without Adaptation}

Using differential series renders it possible to design forwardstepping control that is convergent without the need of adaptation. In particular, we consider a differential-cascaded system yielded by the control (\ref{eq:4}) with the adaptation law and feedforward compensation being removed and the reference dynamics being given as (\ref{eq:16}), namely
\be
\label{eq:20}
\begin{cases}
\frac{d^{\ell+1} \Delta q}{dt^{\ell+1}}=-\alpha_\ell \frac{d^\ell \Delta q}{dt^\ell}-\alpha_{\ell-1} \frac{d^{\ell-1} \Delta q}{dt^{\ell-1}}-\dots-\alpha_0\Delta q+\frac{d^\ell s}{dt^\ell}\\
M(q)\ddot q+C(q,\dot q)\dot q+g(q)=-K s+\tau^\ast.
\end{cases}
\ee
 The control action around the input of the system can simply be written as
 \be
 \label{eq:21}
 \tau=-K(\dot q-z).
  \ee
  The remainder can be given as $R_t=- K^{-1}\frac{d^{\ell}}{dt^{\ell}}[M(q)\ddot q+C(q,\dot q)\dot q+g(q)-\tau^\ast]$. To have a new perspective concerning the standard PID control, we introduce another reference dynamics as
 \be
 \label{eq:22}
K_D\dot z=K_D\ddot q_d-{K}_P \Delta \dot q-{ K}_I \Delta q
 \ee
 where $K_D$, $K_P$, and $K_I$ are the derivative, proportional, and integral gain matrices (symmetric and positive definite), respectively. With the control torque being given as
 \be
 \label{eq:23}
 \tau=-K_D (\dot q-z),
 \ee
we have a differential-cascaded system as
\be
\label{eq:24}
\begin{cases}
K_D \Delta \ddot q=-{K}_P\Delta \dot q-{ K}_I\Delta q +K_D\dot s\\
M(q)\ddot q+C(q,\dot q)\dot q+g(q)-\tau^\ast=-K_D s.
\end{cases}
\ee
We can combine the two subsystems of (\ref{eq:24}) as
\begin{align}
\label{eq:25}
&\frac{d}{dt}\left[M(q)\ddot q+C(q,\dot q)\dot q+g(q)-\tau^\ast\right]\nn\\
&=-K_D\Delta \ddot q-K_P \Delta \dot q-K_I\Delta q.
\end{align}
We note that the differentiation of $M(q)\ddot q+C(q,\dot q)\dot q+g(q)-\tau^\ast$ is involved, which is a result due to the differential-cascaded structure. From (\ref{eq:24}), the remainder can be written as
\begin{align}
R_t=-\frac{d}{dt}\left[M(q)\ddot q+C(q,\dot q)\dot q+g(q)-\tau^\ast\right].
\end{align}
The comparison between (\ref{eq:20}) and (\ref{eq:24}) shows that the standard PID control yields a special form of a general differential-cascaded structure with order $\ell$.

\emph{Theorem 2:} Suppose that the disturbance torque $\tau^\ast$ and its derivative of arbitrary order are bounded and that the remainder
\be
R_t=-K^{-1}\frac{d^{\ell}}{dt^{\ell}}\left[M(q)\ddot q+C(q,\dot q)\dot q+g(q)-\tau^\ast\right]\to 0
\ee
as $\ell\to \infty$. Then, the forwardstepping controller given as (\ref{eq:16}) and (\ref{eq:21}) ensures the convergence of the joint tracking errors, i.e., $\Delta q\to 0$ and $\Delta \dot q\to 0$ as $\ell\to\infty$ and $t\to\infty$.

The forwardstepping control without relying on the dynamics of robot manipulators typically requires the gains to be relatively high or the desired trajectory to be relatively slow. The incorporation of the feedforward control, as is known, is beneficial for permitting the specification of low gains and thus yields strong robustness of the system. In the case that the parameter $\vartheta$ is known, the control torque given by (\ref{eq:4}) becomes
\be
\label{eq:28}
\tau=-K s+Y(q,\dot q,z,\dot z)\vartheta,
\ee
and the reference dynamics are redefined as
\begin{align}
\label{eq:29}
\frac{d^\ell z}{dt^\ell}=&\frac{d^{\ell+1} q_d}{dt^{\ell+1}}-\alpha_\ell \frac{d^\ell \Delta q}{dt^\ell}-\alpha_{\ell-1} \frac{d^{\ell-1} \Delta q}{dt^{\ell-1}}-\dots-\alpha_0\Delta q\nn\\
&+\Lambda \frac{d^{\ell-1}}{dt^{\ell-1}}(\dot q-z)
\end{align}
where $\Lambda$ is a symmetric positive definite matrix. The dynamics of the system can be formulated as the following differential-cascaded system
\be
\label{eq:30}
\begin{cases}
\frac{d^{\ell+1} \Delta q}{dt^{\ell+1}}=-\alpha_\ell \frac{d^\ell\Delta q}{dt^\ell}-\alpha_{\ell-1} \frac{d^{\ell-1} \Delta q}{dt^{\ell-1}}-\dots-\alpha_0\Delta q\\
\quad+\Lambda \frac{d^{\ell-1}s}{dt^{\ell-1}}+\frac{d^\ell s}{dt^\ell}\\
M(q)\dot s+C(q,\dot q)s=-K s+\tau^\ast.
\end{cases}
\ee
Via (\ref{eq:30}), we have that
\begin{align}
\label{eq:31}
\frac{d^{\ell+1} \Delta q}{dt^{\ell+1}}=&-\alpha_\ell \frac{d^\ell \Delta q}{dt^\ell}-\alpha_{\ell-1} \frac{d^{\ell-1} \Delta q}{dt^{\ell-1}}-\dots-\alpha_0\Delta q\nn\\
&+\underbrace{\frac{d^\ell s}{dt^\ell}-\Lambda K^{-1}\frac{d^{\ell-1}}{dt^{\ell-1}}[M(q)\dot s+C(q,\dot q)s-\tau^\ast]}_{R_t}
\end{align}
where $R_t$ is the remainder, mainly due to the nonlinearity of $M(q)$.

\emph{Theorem 3:} Suppose that the disturbance torque $\tau^\ast$ and its derivative of arbitrary order are bounded and that the remainder
\be
R_t=\frac{d^\ell s}{dt^\ell}-\Lambda K^{-1}\frac{d^{\ell-1}}{dt^{\ell-1}}[M(q)\dot s+C(q,\dot q)s-\tau^\ast]\to 0
\ee
as $\ell\to\infty$. Then, the forwardstepping controller given as (\ref{eq:28}) and (\ref{eq:29}) ensures the convergence of the joint tracking errors as $\ell\to\infty$.

In the presence of parametric uncertainty, the feedforward control action can no longer be precisely exerted, which yields the consequence that the stability of the closed-loop dynamics around the input cannot be guaranteed. Suppose that $\hat\vartheta$ is an a priori estimate of $\vartheta$, and we consider the following control with the incorporation of the standard nonlinear damping (see, e.g., \cite{Krstic1995_TAC})
\begin{align}
\label{eq:33}
\tau=&-K s+Y(q,\dot q,z,\dot z)\hat\vartheta\nn\\
&-\lambda_D Y(q,\dot q,z,\dot z)Y^T(q,\dot q,z,\dot z)s
\end{align}
where $\lambda_D$ is a positive design constant. The closed-loop dynamics around the input can be written as
\begin{align}
M&(q)\dot s+C(q,\dot q)s\nn\\
=&-Ks+Y(q,\dot q,z,\dot z)\Delta\vartheta\nn\\
&-\lambda_DY(q,\dot q,z,\dot z)Y^T(q,\dot q,z,\dot z)s+\tau^\ast,
\end{align}
upon which we have from the typical practice that $s\in\mathcal L_\infty$.

\emph{Theorem 4:} Suppose that the disturbance torque $\tau^\ast$ and its derivative of arbitrary order are bounded and that the remainder
$
R_t=\frac{d^\ell s}{dt^\ell}-\Lambda K^{-1}\frac{d^{\ell-1}}{dt^{\ell-1}}[M(q)\dot s+C(q,\dot q)s-Y\Delta\vartheta+\lambda_D YY^Ts-\tau^\ast]\to 0
$
as $\ell\to\infty$. Then, the forwardstepping controller with nonlinear damping given as (\ref{eq:33}) and (\ref{eq:29}) ensures the convergence of the joint tracking errors as $\ell\to\infty$.

As is observed and also well known, the nonlinear damping action belongs to the high-gain feedback, and hence the robustness of the closed-loop dynamics relies closely on the magnitude of $\lambda_D$, yielding the result that $\lambda_D$ is typically expected to be relatively low. To alleviate this issue, we introduce the following dynamics (extending the standard passive filter \cite{Berghuis1993_SCL,Kelly1993_IFAC})
\be
\label{eq:35}
\dot \xi+\lambda \xi=\lambda Y^T(q,\dot q,z,\dot z) s
\ee
where $\lambda$ is a positive design constant. The torque control is specified as
\be
\label{eq:36}
\tau=-K s+Y(q,\dot q,z,\dot z)\hat\vartheta-\lambda_D Y(q,\dot q,z,\dot z)\xi,
\ee
and the closed-loop dynamics around the input become
\begin{align}
M(q)\dot s+C(q,\dot q)s=&-Ks+Y(q,\dot q,z,\dot z)\Delta\vartheta\nn\\
&-\lambda_DY(q,\dot q,z,\dot z)\xi+\tau^\ast.
\end{align}
We now analyze the input-output properties of the system with $\Delta\vartheta$ as the input and with $s$ and $\xi$ as the output, which differs from the typical context using nonlinear damping where such input-output properties are formulated (see, e.g., \cite{Krstic1995_TAC}). We consider the function
\be
\label{eq:38}
V=\frac{1}{2}s^TM(q)s+\frac{\lambda_D}{2\lambda}\xi^T \xi
\ee
and using Property 2 and the standard result concerning basic inequalities, we have that
\begin{align}
\label{eq:39}
\dot V=&-s^T K s-\lambda_D \xi^T \xi+[(1/\lambda)\dot \xi+\xi]^T\Delta\vartheta+s^T\tau^\ast\nn\\
\le&-(1/2)s^T K s-(\lambda_D/2)\xi^T \xi+(1/\lambda)\dot \xi^T\Delta\vartheta\nn\\
&+(1/2)\tau^{\ast T} K^{-1}\tau^\ast+[1/(2\lambda_D)]\Delta\vartheta^T\Delta\vartheta.
\end{align}
Integrating (\ref{eq:39}) and using the standard integral by parts yields
\begin{align}
V&(t)-V(0)\nn\\
\le& -(1/2)\int_0^t s^T K sd\sigma-(\lambda_D/2)\int_0^t \xi^T\xi d\sigma\nn\\
&+(1/\lambda)[\xi^T\Delta\vartheta-\xi^T(0)\Delta\vartheta(0)]\nn\\
&-(1/\lambda)\int_0^t \xi^T\dot{\hat\vartheta}d\sigma+(1/2)\int_0^t \tau^{\ast T} K^{-1}\tau^\ast d\sigma\nn\\
&+[1/(2\lambda_D)]\int_0^t \Delta\vartheta^T\Delta\vartheta d\sigma
\end{align}
and via the application of the standard basic inequalities, we can further have that
\begin{align}
\label{eq:41}
V&(t)-V(0)\nn\\
\le&-(1/2)\int_0^t s^T K sd\sigma-(\lambda_D/4)\int_0^t \xi^T\xi d\sigma\nn\\
&+[\lambda_D/(4\lambda)]\xi^T\xi+[1/(\lambda\lambda_D)]\Delta\vartheta^T\Delta\vartheta\nn\\
&-(1/\lambda)\xi^T(0)\Delta\vartheta(0)+[1/(\lambda^2\lambda_D)]\int_0^t\dot{\hat\vartheta}^T\dot{\hat\vartheta}d\sigma\nn\\
&+(1/2)\int_0^t \tau^{\ast T} K^{-1}\tau^\ast d\sigma\nn\\
&+[1/(2\lambda_D)]\int_0^t \Delta\vartheta^T\Delta\vartheta d\sigma.
\end{align}
Hence, we have from (\ref{eq:38}) and (\ref{eq:41}) that
\begin{align}
\label{eq:42}
V^\ast\le&-(1/2)\int_0^t s^T K sd\sigma-(\lambda_D/4)\int_0^t \xi^T\xi d\sigma\nn\\
&+[1/(2\lambda_D)]\int_0^t\Delta\vartheta^T\Delta\vartheta d\sigma\nn\\
&+(1/2)\int_0^t\tau^{\ast T} K^{-1}\tau^\ast d\sigma+c^\ast
\end{align}
where $V^\ast=(1/2)s^TM(q)s+[\lambda_D/(4\lambda)]\xi^T\xi$, and $c^\ast=V(0)-(1/\lambda)\xi^T(0)\Delta\vartheta(0)+[1/(\lambda\lambda_D)]\Delta\vartheta^T\Delta\vartheta++[1/(\lambda^2\lambda_D)]\int_0^t\dot{\hat\vartheta}^T\dot{\hat\vartheta}d\sigma$. In the case that $\hat\vartheta\in\mathcal L_\infty$ and $\dot{\hat\vartheta}\in\mathcal L_2$, $c^\ast\in\mathcal L_\infty$. Let \be
\gamma^\ast=\min\{\lambda_{\min}\{K\}/\lambda_{\max}\{M(q)\}, \lambda\}
\ee
where $\lambda_{\max}\{\cdot\}$ and $\lambda_{\min}\{\cdot\}$ denote the maximum and minimum eigenvalues of a matrix, respectively. Then, we have an integral inequality from (\ref{eq:42}) as
\be
V^\ast\le -\gamma^\ast \int_0^t [V^\ast(\sigma)+h^\ast(\sigma)]d\sigma+c^\ast
\ee
with $h^\ast=-[1/(2\gamma^\ast\lambda_D)]\Delta\vartheta^T\Delta\vartheta-[1/(2\gamma^\ast)]\tau^{\ast T}K^{-1}\tau^\ast\in\mathcal L_\infty$, and the above inequality can further be written as
\be
V^\ast+h^\ast\le -\gamma^\ast \int_0^t [V^\ast(\sigma)+h^\ast(\sigma)]d\sigma+c^\ast+h^\ast,
\ee
which directly yields the result that $V^\ast+h^\ast\in\mathcal L_\infty$ via the standard practice. Due to the result that $h^\ast\in\mathcal L_\infty$, we have that $V^\ast\in\mathcal L_\infty$, and hence $s\in\mathcal L_\infty$ and $\xi\in\mathcal L_\infty$. In the case that $\hat\vartheta\in\mathcal L_\infty$ and $\dot{\hat\vartheta}\in\mathcal L_\infty$, it can be shown with a similar procedure that the same conclusion holds.

\emph{Theorem 5:} Suppose that the disturbance torque $\tau^\ast$ and its derivative of arbitrary order are bounded and that the remainder
$
R_t=\frac{d^\ell s}{dt^\ell}-\Lambda K^{-1}\frac{d^{\ell-1}}{dt^{\ell-1}}[M(q)\dot s+C(q,\dot q)s-Y\Delta\vartheta+\lambda_D Y\xi-\tau^\ast]\to 0
$
as $\ell\to\infty$. Then, the forwardstepping controller given as (\ref{eq:35}), (\ref{eq:36}), and (\ref{eq:29}) ensures the convergence of the joint tracking errors as $\ell\to\infty$.

\emph{Remark 3:} The formulation of PID control in a differential-cascaded context, differing from the typical perspective (e.g., the perspective of linear systems with respect to the integral of the error), highlights the connection between the structure of the reformulated PID control and the general differential-cascaded framework. The closed-loop dynamics under PID control, via reformulation, can be classified as a most basic form of differential-cascaded structures, namely a first-order differential-cascaded structure. It is well recognized that PID control is confined to accommodating constant or slowly time-varying disturbances or desired trajectories and is typically rigorously analyzed in the context of linear time-invariant systems; the rigorous analysis of PID control for nonlinear systems such as robot manipulators generally relies on the model of the system and involves a relatively complicated condition for ensuring the stability (see, e.g., \cite{Cheah1999_ICRA}). The formulated general differential-cascaded approach here provides a constructive way to handle highly time-varying disturbances and desired trajectories while the condition for rigorously guaranteeing the stability is quite moderate, without involving the uncertain physical parameters of robot manipulators, as is shown in Theorem 4 and Theorem 5.

\subsection{Forwardstepping Control with Forwardstepping Adaptation}

The standard adaptation in the presence of disturbances becomes fragile due to the robustness issue, which yields the challenging implementation of adaptive control in practical applications. Here, we present new adaptation via exploiting symmetry of the differential-cascaded structure, referred to as forwardstepping adaptation. The differential-cascaded structure yields the result that both the order of the target dynamics and that of the closed-loop dynamics around the input are increased and that the elements of target dynamics and those of the closed-loop dynamics around the input can be transferred across the differential-cascaded interconnection.

We consider the same torque control as that given as (\ref{eq:35}) and (\ref{eq:36}) with the adaptation law for $\hat\vartheta$ unspecified. This yields
\be
\label{eq:46}
\begin{cases}
\frac{d^{\ell+1} \Delta q}{dt^{\ell+1}}=-\alpha_\ell \frac{d^\ell\Delta q}{dt^\ell}-\alpha_{\ell-1} \frac{d^{\ell-1} \Delta q}{dt^{\ell-1}}-\dots-\alpha_0\Delta q\\
\quad+\Lambda \frac{d^{\ell-1}s}{dt^{\ell-1}}+\frac{d^\ell s}{dt^\ell}\\
M(q)\dot s+C(q,\dot q)s=-K s+Y(q,\dot q,z,\dot z)\Delta\vartheta\\
\quad-\lambda_DY(q,\dot q,z,\dot z)\xi+\tau^\ast,
\end{cases}
\ee
upon which we have that
\begin{align}
\label{eq:47}
\frac{d^{\ell+1} \Delta q}{dt^{\ell+1}}=&-\alpha_\ell \frac{d^\ell \Delta q}{dt^\ell}-\alpha_{\ell-1} \frac{d^{\ell-1} \Delta q}{dt^{\ell-1}}-\dots-\alpha_0\Delta q\nn\\
&+\Lambda (p I_n+K)^{-1}\frac{d^{\ell-1}}{dt^{\ell-1}}[Y(q,\dot q,z,\dot z)\Delta\vartheta]+R_t
\end{align}
where $p$ denotes the Laplace variable, $I_n$ is the $n\times n$ identity matrix, and $R_t$ is the remainder at the target dynamics given as
\begin{align}
R_t=&\frac{d^\ell s}{dt^\ell}-\Lambda (pI_n+K)^{-1}\frac{d^{\ell-1}}{dt^{\ell-1}}[M(q)\dot s-\dot s+C(q,\dot q)s\nn\\
&+\lambda_D Y\xi-\tau^\ast].
\end{align}
Equation (\ref{eq:47}) can be rewritten as
\begin{align}
\Delta \ddot q+\alpha_1^\ast\Delta \dot q+\alpha_0^\ast \Delta q=&\frac{p^{\ell+1}+\alpha_1^\ast p^\ell+\alpha_0^\ast p^{\ell-1}}{p^{\ell+1}+\alpha_\ell p^\ell+\dots+\alpha_0}\nn\\
&\times\Lambda (pI_n+K)^{-1}[ Y(q,\dot q,z,\dot z)\Delta\vartheta]\nn\\
&+\frac{p^2+\alpha_1^\ast p+\alpha_0^\ast}{p^{\ell+1}+\alpha_\ell p^\ell+\dots+\alpha_0}R_t
\end{align}
where $\alpha_0^\ast$ and $\alpha_1^\ast$ are positive design constants, and without considering the remainder, the mapping from $Y(q,\dot q,z,\dot z)\Delta\vartheta$ to $\Delta \ddot q+\alpha_1^\ast\Delta \dot q+\alpha_0^\ast \Delta q$ is a stable linear system with relative degree one. Let $W$ be defined as
\be
\label{eq:50}
W=\underbrace{\frac{p^{\ell+1}+\alpha_1^\ast p^\ell+\alpha_0^\ast p^{\ell-1}}{p^{\ell+1}+\alpha_\ell p^\ell+\dots+\alpha_0}\Lambda (pI_n+K)^{-1}}_{G(p)} Y(q,\dot q,z,\dot z).
\ee
Due to the fact that $G(p)$ is not strictly positive real as $\ell>1$ (which can be evaluated from \cite{Slotine1991_Book}), we follow the error augmentation approach in \cite{Slotine1991_Book} to develop an adaptation law for $\hat\vartheta$. In particular, define a vector
\be
h=[G(p) Y(q,\dot q,z,\dot z)]\hat\vartheta-G(p)[Y(q,\dot q,z,\dot z)\hat\vartheta]
\ee
and exploiting the fact that
\be
[G(p) Y(q,\dot q,z,\dot z)]\vartheta-G(p)[Y(q,\dot q,z,\dot z)\vartheta]=0,
\ee
we have that
\be
h=[G(p) Y(q,\dot q,z,\dot z)]\Delta \vartheta-G(p)[Y(q,\dot q,z,\dot z)\Delta\vartheta].
\ee
We then have the dynamics concerning $\Delta \ddot q+\alpha_1^\ast\Delta \dot q+\alpha_0^\ast \Delta q+h$ as (without considering the remainder)
\begin{align}
\Delta \ddot q+\alpha_1^\ast\Delta \dot q+\alpha_0^\ast \Delta q+h=W\Delta\vartheta.
\end{align}
The standard practice as in \cite{Slotine1991_Book} indicates the following adaptation law
\be
\dot{\hat\vartheta}=-\Gamma W^T (\Delta \ddot q+\alpha_1^\ast\Delta \dot q+\alpha_0^\ast \Delta q+h),
\ee
yet this only ensures that $\Delta \ddot q+\alpha_1^\ast\Delta \dot q+\alpha_0^\ast \Delta q+h\in\mathcal L_2$ (and $\hat\vartheta\in\mathcal L_\infty$) rather than $\Delta q\in\mathcal L_2$. One possible solution is to redefine the reference dynamics (\ref{eq:29}) with $q_d$ being replaced with $q_d^\ast$ as
\begin{align}
\label{eq:a1}
\frac{d^\ell z}{dt^\ell}=&\frac{d^{\ell+1} q_d^\ast}{dt^{\ell+1}}-\alpha_\ell \frac{d^\ell \Delta q^\ast}{dt^\ell}-\alpha_{\ell-1} \frac{d^{\ell-1} \Delta q^\ast}{dt^{\ell-1}}\nn\\
&-\dots-\alpha_0\Delta q^\ast+\Lambda \frac{d^{\ell-1}}{dt^{\ell-1}}(\dot q-z)
\end{align}
where $\Delta q^\ast=q-q_d^\ast$ and $q_d^\ast$ is specified to be governed by
\begin{align}
\label{eq:56}
&\ddot q_d^\ast+\alpha_1^\ast\dot q_d^\ast+\alpha_0^\ast q_d^\ast-h\nn\\
&=\ddot q_d+\alpha_1^\ast\dot q_d+\alpha_0^\ast q_d-\lambda_D^\ast WW^T(\Delta \dot q+\alpha^\ast\Delta q)
\end{align}
 such that with $\alpha_0^\ast=\alpha^{\ast 2}$ and $\alpha_1^\ast=2\alpha^\ast$
\begin{align}
\label{eq:57}
\frac{d}{dt}(\Delta \dot q+\alpha^\ast\Delta q)=&-\alpha^\ast (\Delta \dot q+\alpha^\ast\Delta q)\nn\\
&-\lambda_D^\ast WW^T(\Delta \dot q+\alpha^\ast\Delta q)\nn\\
&+W\Delta\vartheta
\end{align}
where $\alpha^\ast$ and $\lambda_D^\ast$ are positive design constants. The adaptation law is given as
\be
\label{eq:58}
\dot{\hat\vartheta}=-\Gamma W^T (\Delta \dot q+\alpha^\ast\Delta q).
\ee
The regressor matrix $W$, $\Delta q$, and $\Delta \dot q$, due to the high-order differential-cascaded structure, are no longer strongly influenced by the disturbance $\tau^\ast$. In addition, the dynamics (\ref{eq:57}) with the adaptation law (\ref{eq:58}) are compatible with the input-output properties of the system under the controller (\ref{eq:35}) and (\ref{eq:36}), yielding the stability of the combined design. To implement the forwardstepping adaptive control, via degree-reduction implementation, we only need to calculate $h$.

\emph{Theorem 6:} Suppose that the disturbance torque $\tau^\ast$ and its derivative of arbitrary order are bounded and that the remainder $
R_t=\frac{d^\ell s}{dt^\ell}-\Lambda (pI_n+K)^{-1}\frac{d^{\ell-1}}{dt^{\ell-1}}[M(q)\dot s-\dot s+C(q,\dot q)s+\lambda_D Y\xi-\tau^\ast]\to 0
$ as $\ell\to\infty$. Then, the forwardstepping adaptive controller given as (\ref{eq:35}), (\ref{eq:36}), (\ref{eq:58}), and (\ref{eq:a1}) with $q_d^\ast$ governed by (\ref{eq:56}) ensures the convergence of the joint tracking errors as $\ell\to\infty$.

\emph{Proof:} The dynamics of the system can be described by the following differential-cascaded system
\be
\label{eq:59}
\begin{cases}
\frac{d^{\ell+1} \Delta q^\ast}{dt^{\ell+1}}=-\alpha_\ell \frac{d^\ell\Delta q^\ast}{dt^\ell}-\alpha_{\ell-1} \frac{d^{\ell-1} \Delta q^\ast}{dt^{\ell-1}}-\cdots\\
\quad-\alpha_0\Delta q^\ast+\Lambda \frac{d^{\ell-1}s}{dt^{\ell-1}}+\frac{d^\ell s}{dt^\ell}\\
M(q)\dot s+C(q,\dot q)s=-K s+Y(q,\dot q,z,\dot z)\Delta\vartheta\\
\quad-\lambda_D Y(q,\dot q,z,\dot z)\xi+\tau^\ast\\
\dot{\hat\vartheta}=-\Gamma W^T(\Delta \dot q+\alpha^\ast\Delta q),
\end{cases}
\ee
upon which we have that
\begin{align}
\Delta \ddot q^\ast+\alpha_1^\ast\Delta \dot q^\ast+\alpha_0^\ast \Delta q^\ast+h=W\Delta\vartheta+G^\ast(p) R_t
\end{align}
with $G^\ast(p)=\frac{p^2+\alpha_1^\ast p+\alpha_0^\ast }{p^{\ell+1}+\alpha_\ell p^\ell+\dots+\alpha_0}$, and further that
\begin{align}
\frac{d}{dt}(\Delta \dot q+\alpha^\ast\Delta q)=&-\alpha^\ast (\Delta \dot q+\alpha^\ast\Delta q)\nn\\
&-\lambda_D^\ast WW^T(\Delta \dot q+\alpha^\ast\Delta q)\nn\\
&+W\Delta\vartheta+G^\ast(p) R_t.
\end{align}
Since $R_t\to 0$ as $\ell\to \infty$, we have
\begin{align}
\label{eq:62}
\frac{d}{dt}(\Delta \dot q+\alpha^\ast\Delta q)=&-\alpha^\ast (\Delta \dot q+\alpha^\ast\Delta q)\nn\\
&-\lambda_D^\ast WW^T(\Delta \dot q+\alpha^\ast\Delta q)\nn\\
&+W\Delta\vartheta.
\end{align}
As in the standard practice, we consider the Lyapunov-like function candidate $V=(1/2)(\Delta \dot q+\alpha^\ast\Delta q)^T(\Delta \dot q+\alpha^\ast\Delta q)+(1/2)\Delta\vartheta^T\Gamma^{-1}\Delta\vartheta$, and the derivative of $V$ along the trajectories of (\ref{eq:62}) and (\ref{eq:58}) can be formulated as $\dot V=-\alpha^\ast(\Delta \dot q+\alpha^\ast\Delta q)^T(\Delta \dot q+\alpha^\ast\Delta q)-\lambda_D^\ast(\Delta \dot q+\alpha^\ast\Delta q)^T WW^T(\Delta \dot q+\alpha^\ast\Delta q)\le0$. This implies that $\Delta \dot q+\alpha^\ast\Delta q\in\mathcal L_2\cap\mathcal L_\infty$, $\hat\vartheta\in\mathcal L_\infty$, and $\dot{\hat\vartheta}\in\mathcal L_2$. From the result that $\Delta \dot q+\alpha^\ast\Delta q\in\mathcal L_2\cap\mathcal L_\infty$ and using the input-output properties of exponentially stable and strictly proper linear systems \cite[p.~59]{Desoer1975_Book}, we have that $\Delta q\in\mathcal L_2\cap\mathcal L_\infty$, $\Delta \dot q\in\mathcal L_2\cap\mathcal L_\infty$, and $\Delta q\to 0$ as $t\to\infty$. Hence, $q\in\mathcal L_\infty$ and $\dot q\in\mathcal L_\infty$. From the input-output properties of the second subsystem of (\ref{eq:59}) with $\Delta\vartheta$ as the input and $s$ and $\xi$ as the output which are shown in Sec. 4.2, we have that $s\in\mathcal L_\infty$ and $\xi\in\mathcal L_\infty$. This implies that $z=\dot q-s\in\mathcal L_\infty$ and further that $\dot z$ is integral bounded. From (\ref{eq:50}) and using the input-output properties of exponentially stable linear systems (see, e.g., \cite[p.~59]{Desoer1975_Book} and \cite{Wang2020_AUT}), we have that $W$ is bounded due to the fact that $Y(q,\dot q,z,\dot z)$ can be considered to be the sum of a bounded quantity and an integral-bounded quantity. $Y(q,\dot q,z,\dot z)\hat\vartheta$ can be written as the sum of bounded, integral-bounded, and square-integrable variables, and using the input-output properties of exponentially stable and strictly proper linear systems in \cite[p.~59]{Desoer1975_Book} and \cite{Wang2020_AUT} yields the result that $G(p)[Y(q,\dot q,z,\dot z)\hat\vartheta]\in\mathcal L_\infty$. Therefore, $h\in\mathcal L_\infty$, and from (\ref{eq:56}), we have that $q_d^\ast\in\mathcal L_\infty$, $\dot q_d^\ast\in\mathcal L_\infty$, and $\ddot q_d^\ast\in\mathcal L_\infty$ based on the input-output properties of exponentially stable linear systems \cite[p.~59]{Desoer1975_Book}. Then, we have from (\ref{eq:a1}) that $\dot z\in\mathcal L_\infty$ and further that $\ddot q\in\mathcal L_\infty$ using similar procedures as in the proof of Theorem 1. This implies that $\Delta \ddot q\in\mathcal L_\infty$ and further that $\Delta \dot q$ is uniformly continuous. The application of Barbalat's lemma \cite{Slotine1991_Book} yields the conclusion that $\Delta \dot q\to0$ as $t\to\infty$. \hfill\text{\small $\blacksquare$}

\emph{Remark 4:} For a desired trajectory or disturbance torque containing periodical information with the frequencies being higher than one [for instance, $\sin(2t)$], the remainder $R_t$ no longer converges to zero. However, the first subsystem of (\ref{eq:59}) contains an inherent time scaling effect, and using a design constant $\kappa>1$, we have that
\begin{align}
\label{eq:63}
\kappa \frac{d^{\ell+1} \Delta q^\ast}{d(\kappa t)^{\ell+1}}=&-\alpha_\ell \frac{d^\ell \Delta q^\ast}{d(\kappa t)^\ell}-\frac{\alpha_{\ell-1}}{\kappa} \frac{d^{\ell-1} \Delta q^\ast}{d(\kappa t)^{\ell-1}}-\dots-\frac{\alpha_0}{\kappa^\ell}\Delta q^\ast\nn\\
&+\frac{\Lambda (p I_n+K)^{-1}}{\kappa}\frac{d^{\ell-1}}{d(\kappa t)^{\ell-1}}[Y(q,\dot q,z,\dot z)\Delta\vartheta]\nn\\
&+\frac{R_t}{\kappa^\ell}.
\end{align}
If we specify $\kappa$ to be such that $\kappa>\omega_{\max}$ with $\omega_{\max}$ denoting the maximum frequency, then the remainder $R_t/\kappa^\ell\to 0$ as $\ell\to \infty$, and at the same time, we expect that the gain $\alpha_0/\kappa^\ell$ (which approximately quantifies the performance at the steady state) does not converge to zero as $\ell\to\infty$. Hence, the time scale $\kappa$ is mainly determined by $\alpha_0$. For ensuring the asymptotic stability of the reformulated polynomial, $\alpha_0,\dots,\alpha_\ell$ can be chosen such that $ w^{\ell+1}+(\alpha_\ell/\kappa)w^\ell+[\alpha_{\ell-1}/\kappa^2]w^{\ell-1}+\dots+\alpha_0/\kappa^{\ell+1}$ is a Hurwitz polynomial. On the other hand, if we specify $\alpha_0,\dots,\alpha_\ell$ such that the polynomial $w^{\ell+1}+\alpha_\ell w^\ell+\dots+\alpha_0$ is a critically-damped stable one, the time-scaling effect no longer depends on the redefinition of $\alpha_0,\dots,\alpha_\ell$. In fact, in the critically-damped case, $\alpha_0=\alpha^{\ell+1},\dots,\alpha_\ell=(\ell+1)\alpha$ with $\alpha$ being a specified constant a priori, and with the time-scaling $\kappa$, $\alpha_0/(\kappa^{\ell+1}),\dots,\alpha_\ell/\kappa$ still yield a critically-damped polynomial with the a priori specified constant $\alpha$ being changed to $\alpha/\kappa$. The consequence due to the presence of high-frequency information is that the speed of convergence becomes slow and the performance is decreased. In addition, the forwardstepping adaptive control in Sec. 3 can be directly modified to yield reliable adaptation in the presence of sustaining disturbances, namely using (\ref{eq:35}), (\ref{eq:36}), and (\ref{eq:58}) to define the control torque.

\emph{Remark 5:} The presence of a remainder is the result of guaranteeing both the adaptability and robustness of robot manipulators, which are typically considered to be conflicting in the literature (see, e.g., \cite{Ioannou1996_Book,Slotine1991_Book}). This is mainly due to the constraint of the application of the conventional order-reduction paradigm in which case it is hard or even impossible to simultaneously achieve the two objectives. Using an order-increment paradigm with infinite differential series, this long-standing problem becomes tractable with the adaptation and robustness being accommodated in a separate way. Even in the case of a constant unknown inertia or inertia matrix (for instance, forwardstepping adaptive control of a point mass with disturbances), it may still not be possible to completely remove the remainder.

\section{Forwardstepping Control with Stacked Reference Dynamics}

The introduction of a single reference dynamic system (reference dynamics) exhibits limited capability of handling large-frequency part of the disturbance or desired trajectory. To accommodate this problem efficiently (for instance, with a low-order differential-cascaded structure), we specify multiple reference dynamics in a stacked way, and the large-frequency signal is compensated via specifying an adaptive differential-cascaded structure. Intuitively, for a periodical signal $\sin(\omega t)$ with $\omega$ being the frequency, it satisfies the standard equation ${d^2\sin(\omega t)}/{dt^2}+\omega^2 \sin(\omega t)=0$, and this equation can be exploited for the design of differential-cascaded structures.

Define a stacked reference dynamic system as
\begin{align}
\label{eq:64}
\frac{d^\ell z}{dt^\ell}=&\frac{d^{\ell+1} q_d^\ast}{dt^{\ell+1}}-\alpha_\ell \frac{d^\ell \Delta q^\ast}{dt^\ell}-\alpha_{\ell-1} \frac{d^{\ell-1} \Delta q^\ast}{dt^{\ell-1}}\nn\\
&-\dots-\alpha_0\Delta q^\ast+\Lambda \frac{d^{\ell-1}}{dt^{\ell-1}}(\dot q-z)\\
\label{eq:65}
\ddot \chi=&\dot z+\hat \theta (q-\chi)
\end{align}
where $\hat\theta$ is the estimate of $\theta$ with $\theta$ denoting the square of the unknown frequency. Define
\begin{align}
\psi=&q-\chi\\
\dot\chi^\ast=&\dot\chi-\alpha^\ast \psi\\
s^\ast=&\dot q-\dot\chi^\ast
\end{align}
where $\alpha^\ast$ is a positive design constant.
The control torque is specified to be similar to the previous result, namely
\be
\label{eq:69}
\tau=-K s^\ast+Y(q,\dot q,\dot\chi^\ast,\ddot \chi^\ast)\hat\vartheta-\lambda_DY(q,\dot q,\dot\chi^\ast,\ddot \chi^\ast)\xi
\ee
with $\xi$ being generated by the dynamics
\be
\label{eq:70}
\dot \xi+\lambda \xi=\lambda Y^T(q,\dot q,\dot\chi^\ast,\ddot \chi^\ast)s^\ast.
\ee
The dynamics of the system can now be described by a stacked differential-cascaded system as
\be
\label{eq:71}
\begin{cases}
\frac{d^{\ell+1} \Delta q^\ast}{dt^{\ell+1}}=-\alpha_\ell \frac{d^\ell\Delta q^\ast}{dt^\ell}-\alpha_{\ell-1} \frac{d^{\ell-1} \Delta q^\ast}{dt^{\ell-1}}-\dots-\alpha_0\Delta q^\ast\\
\quad+\Lambda \frac{d^{\ell-1} s}{dt^{\ell-1}}+\frac{d^\ell s}{dt^\ell}\\
\dot s=\Delta \theta \psi+\theta\psi+\ddot \psi \\
\dot\psi=-\alpha^\ast \psi+s^\ast\\
M(q)\dot s^\ast+C(q,\dot q)s^\ast=-Ks^\ast+Y(q,\dot q,\dot\chi^\ast,\ddot \chi^\ast)\Delta\vartheta\\
\quad-\lambda_D Y(q,\dot q,\dot\chi^\ast,\ddot \chi^\ast)\xi+\tau^\ast
\end{cases}
\ee
where $\Delta \theta=\hat\theta-\theta$. Via the first two subsystems of (\ref{eq:71}), we have that
\begin{align}
\label{eq:72}
\frac{d^{\ell+1} \Delta q^\ast}{dt^{\ell+1}}=&-\alpha_\ell \frac{d^\ell \Delta q^\ast}{dt^\ell}-\alpha_{\ell-1} \frac{d^{\ell-1} \Delta q^\ast}{dt^{\ell-1}}-\dots-\alpha_0\Delta q^\ast\nn\\
&+\Lambda\frac{d^{\ell-2}}{dt^{\ell-2}}(\Delta\theta\psi)+\frac{d^{\ell-1}}{dt^{\ell-1}}(\Delta\theta\psi)\nn\\
&+\theta\left[\Lambda\frac{d^{\ell-2}}{dt^{\ell-2}}\psi+\frac{d^{\ell-1}}{dt^{\ell-1}}\psi\right]\nn\\
&+\Lambda\frac{d^{\ell}}{dt^{\ell}}\psi+\frac{d^{\ell+1}}{dt^{\ell+1}}\psi
\end{align}
and with the combination of (\ref{eq:72}) and the lower two subsystems of (\ref{eq:71}), we have
\begin{align}
\label{eq:73}
\frac{d^{\ell+1} \Delta q^\ast}{dt^{\ell+1}}=&-\alpha_\ell \frac{d^\ell \Delta q^\ast}{dt^\ell}-\alpha_{\ell-1} \frac{d^{\ell-1} \Delta q^\ast}{dt^{\ell-1}}-\dots-\alpha_0\Delta q^\ast\nn\\
&+\Lambda\frac{d^{\ell-2}}{dt^{\ell-2}}(\Delta\theta\psi)+\frac{d^{\ell-1}}{dt^{\ell-1}}(\Delta\theta\psi)\nn\\
&+\theta(p+\alpha^\ast)^{-1}\frac{d^{\ell-2}}{dt^{\ell-2}}\Big\{\dot s^\ast-\Lambda (pI_n+K)^{-1}\nn\\
&\times[M(q)\dot s^\ast-\dot s^\ast+C(q,\dot q)s^\ast-Y\Delta\vartheta\nn\\
&+\lambda_DY\xi-\tau^\ast] \Big\}+(p+\alpha^\ast)^{-1}\frac{d^{\ell}}{dt^{\ell}}\Big\{\dot s^\ast\nn\\
&-\Lambda (pI_n+K)^{-1}[M(q)\dot s^\ast-\dot s^\ast\nn\\
&+C(q,\dot q)s^\ast-Y\Delta\vartheta+\lambda_DY\xi-\tau^\ast] \Big\}.
\end{align}
We can further write (\ref{eq:73}) as
\begin{align}
\label{eq:74}
\frac{d^{\ell+1} \Delta q^\ast}{dt^{\ell+1}}=&-\alpha_\ell \frac{d^\ell \Delta q^\ast}{dt^\ell}-\alpha_{\ell-1} \frac{d^{\ell-1} \Delta q^\ast}{dt^{\ell-1}}-\dots-\alpha_0\Delta q^\ast\nn\\
&+\Lambda\frac{d^{\ell-2}}{dt^{\ell-2}}(\Delta\theta\psi)+\frac{d^{\ell-1}}{dt^{\ell-1}}(\Delta\theta\psi)\nn\\
&+(p+\alpha^\ast)^{-1}\Lambda(pI_n+ K)^{-1}\bigg[\theta\frac{d^{\ell-2}}{dt^{\ell-2}}(Y\Delta\vartheta)\nn\\
&+\frac{d^{\ell}}{dt^{\ell}}(Y\Delta\vartheta)\bigg]+R_t
\end{align}
with
\begin{align}
R_t=&(p+\alpha^\ast)^{-1}\theta\frac{d^{\ell-2}}{dt^{\ell-2}}\Big\{\dot s^\ast-\Lambda (pI_n+K)^{-1}[M(q)\dot s^\ast\nn\\
&-\dot s^\ast+C(q,\dot q)s^\ast+\lambda_DY\xi-\tau^\ast]\Big\}\nn\\
&+(p+\alpha^\ast)^{-1}\frac{d^{\ell}}{dt^{\ell}}\Big\{\dot s^\ast-\Lambda (p I_n+ K)^{-1}[M(q)\dot s^\ast\nn\\
&-\dot s^\ast+C(q,\dot q)s^\ast+\lambda_DY\xi-\tau^\ast]\Big\}.
\end{align}
Equation (\ref{eq:74}) can be rewritten as
\begin{align}
\label{eq:76}
\Delta& \ddot q^\ast+\alpha_1^\ast\Delta \dot q^\ast+\alpha_0^\ast\Delta q^\ast\nn\\
=&G_1(p)(\psi\Delta \theta)+\theta G_2(p)(Y\Delta\vartheta)\nn\\
&+G_3(p)(Y\Delta\vartheta)+G^\ast(p)R_t
\end{align}
where
\begin{align}
G_1(p)=&G^\ast(p)(p^{\ell-1}I_n+p^{\ell-2}\Lambda )\\
G_2(p)=&G^\ast(p)(p+\alpha^\ast)^{-1}p^{\ell-2}\Lambda (pI_n+K)^{-1}\\
G_3(p)=&G^\ast(p)(p+\alpha^\ast)^{-1}p^{\ell}\Lambda (pI_n+K)^{-1}.
\end{align}
Following the previous practice, we define
\begin{align}
h=&[G_1(p)\psi]\hat\theta-G_1(p)(\psi\hat\theta)\nn\\
&+\hat \theta \{[G_2(p)Y]\hat\vartheta-G_2(p)(Y\hat\vartheta)\}\nn\\
&+[G_3(p)Y]\hat\vartheta-G_3(p)(Y\hat\vartheta)
\end{align}
and with the result
\begin{align}
&[G_1(p)\psi]\theta-G_1(p)(\psi\theta)=0\\
&[G_2(p)Y]\vartheta-G_2(p)(Y\vartheta)=0\\
&[G_3(p)Y]\vartheta-G_3(p)(Y\vartheta)=0,
\end{align}
we have that
\begin{align}
\label{eq:84}
h=&[G_1(p)\psi]\Delta\theta-G_1(p)(\psi\Delta\theta)\nn\\
&+\hat\theta\{[G_2(p)Y]\Delta\vartheta-G_2(p)(Y\Delta\vartheta)\}\nn\\
&+[G_3(p)Y]\Delta\vartheta-G_3(p)(Y\Delta\vartheta).
\end{align}
Using (\ref{eq:84}), we can rewrite (\ref{eq:76}) as
\begin{align}
\Delta \ddot q^\ast+\alpha_1^\ast\Delta\dot q^\ast+\alpha_0^\ast \Delta q^\ast+h=&W_1\Delta\theta+W^\ast \Delta\vartheta+R_t^\ast
\end{align}
where $W_1=G_1(p)\psi$, $W^\ast=\big\{ \hat \theta [G_2(p)Y]+G_3(p)Y\big\}$, and $R_t^\ast=-\Delta\theta G_2(p)(Y\Delta\vartheta)+G^\ast(p)R_t$ is the remainder which incorporates the nonlinearly parametric term. Define $q_d^\ast$ using the dynamics
\begin{align}
\ddot q_d^\ast+\alpha_1^\ast \dot q_d^\ast+\alpha_0^\ast q_d^\ast-h=&\ddot q_d+\alpha_1^\ast \dot q_d+\alpha_0^\ast q_d\nn\\
&-\lambda_D^\ast W^\ast W^{\ast T}(\Delta \dot q+\alpha^\ast\Delta q)
\end{align}
and we have that
\begin{align}
\frac{d}{dt}(\Delta \dot q+\alpha^\ast\Delta q)=&-\alpha^\ast (\Delta \dot q+\alpha^\ast\Delta q) \nn\\
&-\lambda_D^\ast W^\ast W^{\ast T}(\Delta \dot q+\alpha^\ast\Delta q)\nn\\
&+W_1\Delta\theta+W^\ast\Delta \vartheta+R_t^\ast.
\end{align}
The adaptation laws for $\hat\theta$ and $\hat\vartheta$ are given as
\begin{align}
\label{eq:88}
\dot{\hat\theta}=&-\gamma W_1^T(\Delta \dot q+\alpha^\ast\Delta q)\\
\label{eq:89}
\dot{\hat\vartheta}=&-\Gamma W^{\ast T}(\Delta \dot q+\alpha^\ast\Delta q)
\end{align}
where $\gamma$ is a positive design constant.

\emph{Theorem 7:} Suppose that the disturbance torque $\tau^\ast$ and its derivative of arbitrary order are bounded and that the remainder $
R_t^\ast\to 0
$ as $\ell\to\infty$. Then, the forwardstepping adaptive controller given as (\ref{eq:64}), (\ref{eq:65}), (\ref{eq:69}), (\ref{eq:70}), (\ref{eq:88}), and (\ref{eq:89}) ensures the convergence of the joint tracking errors as $\ell\to\infty$.

The proof of Theorem 7 can be conducted via following similar practice as previously performed.

\emph{Remark 6:} To handle the case with a disturbance containing signals with two large unknown frequencies $\omega_1$ and $\omega_2$, we can consider the following differential-cascaded system
\be
\begin{cases}
\frac{d^{\ell+1} \Delta q^\ast}{dt^{\ell+1}}=-\alpha_\ell \frac{d^\ell\Delta q^\ast}{dt^\ell}-\alpha_{\ell-1} \frac{d^{\ell-1} \Delta q^\ast}{dt^{\ell-1}}-\dots-\alpha_0\Delta q^\ast\\
\quad+\Lambda \frac{d^{\ell-1} s}{dt^{\ell-1}}+\frac{d^\ell s}{dt^\ell}\\
\dddot s=\hat\theta_1\psi+\hat\theta_2\ddot\psi+\frac{d^4}{dt^4} \psi \\
\dot\psi=-\alpha^\ast \psi+s^\ast\\
M(q)\dot s^\ast+C(q,\dot q)s^\ast=-Ks^\ast+Y(q,\dot q,\dot\chi^\ast,\ddot \chi^\ast)\Delta\vartheta\\
\quad-\lambda_D Y(q,\dot q,\dot\chi^\ast,\ddot \chi^\ast)\xi+\tau^\ast
\end{cases}
\ee
where $\hat\theta_1$ and $\hat\theta_2$ are the estimates of $\theta_1=\omega_1^2\omega_2^2$ and $\theta_2=\omega_1^2+\omega_2^2$, respectively. This shows that in the case of two unknown frequencies, the adaptation needs to be conducted with respect to a group of parameters that are nonlinear functions of the squares of the frequencies, differing from that of one unknown frequency. The case with multiple large unknown frequencies can be accommodated using similar procedures.

\emph{Remark 7:} We here encounter the nonlinear parametrization problem as handling the uncertainty of robot dynamics and frequency uncertainty of the disturbance due to their intertwining nature. This problem is solved via neglecting the high-order terms with respect to the uncertainty, similar to the use of the standard Taylor polynomials to approximate functions. The nonlinear parametrization problem associated with the unknown frequencies of the disturbance is well recognized in the context of the standard internal model approach (see, e.g., \cite{Chen2007_CDC,Lu2019_AUT}).

To resolve the approximate issue of the previous solution, we develop another stacked differential-cascaded structure that is based upon the idea that the effect of the disturbances and that of the parametric uncertainty of the dynamics may be separated in certain sense.
Specifically, for the case of two unknown frequencies, we define a stacked reference dynamic system as
\begin{align}
\label{eq:91}
\ddot \chi_1=&\ddot q_d^\ast-\alpha_1^{\ast\ast}\Delta \dot q^\ast-\alpha_0^{\ast\ast}\Delta q^\ast\\
\label{eq:92}
\frac{d^4\chi_2}{dt^4}=&\frac{d^4\chi_1}{d t^4}-\alpha_3^\ast\dddot \psi_1-\alpha_2^\ast\ddot\psi_1-\alpha_1^\ast\dot
\psi_1-\alpha_0^\ast\psi_1\nn\\
&+H^{\ast-1}(p)\{[H^\ast(p)(q-\chi_2)]\hat\theta_1\}\nn\\
&+H^{\ast-1}(p)\{[H^\ast(p)(\ddot q-\ddot\chi_2)]\hat\theta_2\}\\
\label{eq:93}
\frac{d^\ell z}{dt^\ell}=&\frac{d^{\ell+1} \chi_2}{dt^{\ell+1}}-\alpha_\ell \frac{d^\ell \psi_2}{dt^\ell}-\alpha_{\ell-1} \frac{d^{\ell-1} \psi_2}{dt^{\ell-1}}\nn\\
&-\dots-\alpha_0\psi_2+\Lambda \frac{d^{\ell-1}}{dt^{\ell-1}}(\dot q-z)
\end{align}
where $\alpha_0^{\ast\ast}$ and $\alpha_1^{\ast\ast}$ are positive design constants,
\begin{align}
\psi_1=&q-\chi_1\\
\psi_2=&q-\chi_2,
\end{align}
 and $H^\ast(p)=\frac{p^2+\kappa_1^\ast p+\kappa_0^\ast}{p^4+\alpha_3^\ast p^3+\alpha_2^\ast p^2+\alpha_1^\ast p+\alpha_0^\ast}$ with $\kappa_0^\ast$ and $\kappa_1^\ast$ being positive design constants and $\alpha_0^\ast,\alpha_1^\ast,\alpha_2^\ast,\alpha_3^\ast$ being chosen such that $w^4+\alpha_3^\ast w^3+\alpha_2^\ast w^2+\alpha_1^\ast w+\alpha_0^\ast$ is a Hurwitz polynomial.
This yields the following stacked differential-cascaded system
\be
\label{eq:96}
\begin{cases}
\Delta \ddot q^\ast=-\alpha_1^{\ast\ast}\Delta \dot q^\ast-\alpha_0^{\ast\ast}\Delta q^\ast+\ddot \psi_1\\
\frac{d^4}{dt^4}\psi_1=-\alpha_3^\ast\dddot\psi_1-\alpha_2^\ast \ddot \psi_1-\alpha_1^\ast \dot\psi_1-\alpha_0^\ast\psi_1\\
\quad+H^{\ast-1}(p)\{[H^\ast(p)\psi_2]\Delta\theta_1\}\\
\quad+H^{\ast-1}(p)\{[H^\ast(p)\ddot \psi_2]\Delta\theta_2\}\\
\quad+\theta_1\psi_2+\theta_2 \ddot \psi_2+\frac{d^4}{dt^4} \psi_2 \\
\frac{d^{\ell+1}  \psi_2}{dt^{\ell+1}}=-\alpha_\ell \frac{d^\ell \psi_2}{dt^\ell}-\alpha_{\ell-1} \frac{d^{\ell-1}  \psi_2}{dt^{\ell-1}}-\cdots\\
\quad-\alpha_0 \psi_2+\Lambda \frac{d^{\ell-1} s}{dt^{\ell-1}}+\frac{d^\ell s}{dt^\ell}\\
M(q)\dot s+C(q,\dot q)s=-K s+Y(q,\dot q,z,\dot z)\Delta\vartheta\\
\quad-\lambda_DY(q,\dot q,z,\dot z)\xi+\tau^\ast
\end{cases}
\ee
where $\Delta\theta_i=\hat\theta_i-\theta_i$, $i=1,2$. The distinctive property of this differential-cascaded structure as compared with the previous one is that the state generated by the closed-loop dynamics around the input first passes through a high-order differential-cascaded structure (with order $\ell$), and then two differential-cascaded structures with order zero (namely standard cascaded structures). This renders it possible to handle the uncertainty of the frequencies and that of the robot dynamics in a separate way. In particular, we have from (\ref{eq:96}) that
\begin{align}
\label{eq:97}
\frac{d^{\ell+1} \psi_2}{dt^{\ell+1}}=&-\alpha_\ell \frac{d^\ell \psi_2}{dt^\ell}-\alpha_{\ell-1} \frac{d^{\ell-1} \psi_2}{dt^{\ell-1}}-\dots-\alpha_0\psi_2\nn\\
&+\Lambda (pI_n+K)^{-1}\frac{d^{\ell-1}}{dt^{\ell-1}}[Y(q,\dot q,z,\dot z)\Delta\vartheta]+R_t
\end{align}
where $R_t$ is the remainder given as
$
R_t=\frac{d^\ell s}{dt^\ell}-\Lambda (pI_n+K)^{-1}\frac{d^{\ell-1}}{dt^{\ell-1}}[M(q)\dot s-\dot s+C(q,\dot q)s+\lambda_D Y\xi-\tau^\ast]$. From the first two subsystems of (\ref{eq:96}) and (\ref{eq:97}), we have
\begin{align}
\label{eq:98}
\Delta \ddot q^\ast=&-\alpha_1^{\ast\ast}\Delta \dot q^\ast-\alpha_0^{\ast\ast}\Delta q^\ast\nn\\
&+G_1^\ast(p)H^{\ast-1}(p)[\hat\theta_1H^\ast(p)G_2^\ast(p)\nn\\
&\times\Lambda (pI_n+K)^{-1}(Y\Delta\vartheta)\nn\\
&+\hat\theta_2 p^2H^\ast(p)G_2^\ast(p)\Lambda (pI_n+K)^{-1}(Y\Delta\vartheta)\nn\\
&+p^4H^\ast(p)G_2^\ast(p)\Lambda (pI_n+K)^{-1}(Y\Delta\vartheta)]+R_t^{\ast}
\end{align}
where
\begin{align}
G_1^\ast(p)=&\frac{p^2}{p^4+\alpha_3^\ast p^3+\alpha_2^\ast p^2+\alpha_1^\ast p+\alpha_0^\ast}\\ G_2^\ast(p)=&\frac{p^{\ell-1}}{p^{\ell+1}+\alpha_\ell p^\ell+\dots+\alpha_0}\\
R_t^{\ast}=&G_1^\ast(p)H^{\ast-1}(p)[\hat\theta_1 H^\ast (p)G_2^{\ast\ast}(p)R_t]\nn\\
&+G_1^\ast(p)H^{\ast-1}(p)[\hat\theta_2 p^2H^\ast(p)G_2^{\ast\ast}(p)R_t]\nn\\
&+ G_1^\ast(p)H^{\ast-1}(p)[ p^4 H^\ast(p)G_2^{\ast\ast}(p)R_t]
\end{align} with $G_2^{\ast\ast}(p)=\frac{1}{p^{\ell+1}+\alpha_\ell p^\ell+\dots+\alpha_0}$. Similar to the practice as previously conducted, we define
\begin{align}
h=&\big\{G_1^\ast(p)H^{\ast-1}(p)[\hat\theta_1H^\ast(p)G_2^\ast(p)\Lambda(pI_n+K)^{-1}Y\nn\\
&+\hat\theta_2 p^2H^\ast(p)G_2^\ast(p)\Lambda (p I_n+K)^{-1}Y\nn\\
&+p^4H^\ast(p)G_2^\ast(p)\Lambda (pI_n+K)^{-1}Y]\big\}\hat\vartheta\nn\\
&-G_1^\ast(p)H^{\ast-1}(p)[\hat\theta_1H^\ast(p)G_2^\ast(p)\Lambda(pI_n+K)^{-1}(Y\hat\vartheta)\nn\\
&+\hat\theta_2 p^2H^\ast(p)G_2^\ast(p)\Lambda (pI_n+K)^{-1}(Y\hat\vartheta)\nn\\
&+p^4H^\ast(p)G_2^\ast(p)\Lambda(pI_n+K)^{-1}(Y\hat\vartheta)].
\end{align}
and we have from (\ref{eq:98}) that
\begin{align}
\Delta \ddot q^\ast=&-\alpha_1^{\ast\ast}\Delta \dot q^\ast-\alpha_0^{\ast\ast}\Delta q^\ast-h\nn\\
&+
W^\ast\Delta\vartheta+R_t^{\ast}
\end{align}
where
\begin{align}
W^\ast=&G_1^\ast(p)H^{\ast-1}(p)[\hat\theta_1H^\ast(p)G_2^\ast(p)\Lambda(pI_n+K)^{-1}Y\nn\\
&+\hat\theta_2 p^2H^\ast(p)G_2^\ast(p)\Lambda (pI_n+K)^{-1}Y\nn\\
&+p^4H^\ast(p)G_2^\ast(p)\Lambda(pI_n+ K)^{-1}Y].
\end{align} To facilitate the design of the adaptation law for $\hat\vartheta$, we specify $\alpha_1^{\ast\ast}=2\alpha^{\ast\ast}$ and $\alpha_0^{\ast\ast}=\alpha^{\ast\ast 2}$ with $\alpha^{\ast\ast}$ being a positive design constant, and this yields
\begin{align}
\Delta \ddot q^\ast=&-2\alpha^{\ast\ast}\Delta \dot q^\ast-\alpha^{\ast\ast2}\Delta q^\ast-h\nn\\
&+
W^\ast\Delta\vartheta+R_t^{\ast}.
\end{align}
Define $q_d^\ast$ via the following dynamics
\begin{align}
\ddot q_d^\ast&+2\alpha^{\ast\ast}\dot q_d^\ast+\alpha^{\ast\ast 2}q_d^\ast-h\nn\\
=&\ddot q_d+2\alpha^{\ast\ast}\dot q_d+\alpha^{\ast\ast 2}q_d\nn\\
&-\lambda_D^\ast W^\ast W^{\ast T}(\Delta \dot q+\alpha^{\ast\ast}\Delta q),
\end{align}
and we have that
\begin{align}
\Delta \ddot q=&-2\alpha^{\ast\ast}\Delta \dot q-\alpha^{\ast\ast2}\Delta q-\lambda_D^\ast W^\ast W^{\ast T}(\Delta \dot q+\alpha^{\ast\ast}\Delta q)\nn\\
&+W^\ast\Delta\vartheta+R_t^{\ast}.
\end{align}
The adaptation law for $\hat\vartheta$ is given as
\be
\label{eq:105}
\dot{\hat\vartheta}=-\Gamma W^{\ast T}(\Delta \dot q+\alpha^{\ast\ast}\Delta q).
\ee
For designing the adaptation laws for $\hat\theta_1$ and $\hat\theta_2$, we rewrite the second subsystem of (\ref{eq:96}) as
\begin{align}
\ddot \psi_1=&-\kappa_1^\ast\dot \psi_1-\kappa_0^\ast \psi_1+[H^\ast(p)\psi_2] \Delta \theta_1 +[H^\ast(p)\ddot \psi_2]\Delta\theta_2\nn\\
&+\underbrace{H^\ast(p)\Big(\theta_1\psi_2+\theta_2 \ddot \psi_2+\frac{d^4}{dt^4} \psi_2\Big)}_{R_{\psi_1}}
\end{align}
where $R_{\psi_1}$ is the remainder. Choosing $\kappa_0^\ast$ and $\kappa_1^\ast$ as $\kappa_0^\ast=\kappa^{\ast 2}$ and $\kappa_1^\ast=2\kappa^\ast$, respectively with $\kappa^\ast$ being a positive design constant, we further have that
\begin{align}
\frac{d}{dt} (\dot \psi_1+\kappa^\ast \psi_1)=&-\kappa^\ast(\dot \psi_1+\kappa^\ast \psi_1)\nn\\
&+W_1\Delta\theta_1+W_2\Delta\theta_2+R_{\psi_1}
\end{align}
where $W_1=H^\ast(p)\psi_2$ and $W_2=H^\ast(p)\ddot \psi_2$. The adaptation laws for $\hat\theta_1$ and $\hat\theta_2$ are specified as
\begin{align}
\label{eq:111}
\dot{\hat\theta}_1=-\gamma_1 W_1^T(\dot \psi_1+\kappa^\ast \psi_1)\\
\label{eq:112}
\dot{\hat\theta}_2=-\gamma_2 W_2^T(\dot \psi_1+\kappa^\ast \psi_1)
\end{align}
where $\gamma_1$ and $\gamma_2$ are positive design constants.

\emph{Theorem 8:} Suppose that the disturbance torque $\tau^\ast$ and its derivative of arbitrary order are bounded and that the remainder $
R_t^\ast\to 0
$ as $\ell\to\infty$ and $R_{\psi_1}$ exponentially converges to zero. Then, the forwardstepping adaptive controller given as (\ref{eq:91}), (\ref{eq:92}), (\ref{eq:93}), (\ref{eq:35}), (\ref{eq:36}), (\ref{eq:105}), (\ref{eq:111}), and (\ref{eq:112}) ensures the convergence of the joint tracking errors as $\ell\to\infty$.

The proof of Theorem 8 can be completed with similar procedures as in the previous practice.

In the general case involving a disturbance $\tau^\ast$ with $n^\ast$ unknown frequencies denoted by $\omega_i$, $i=1,\dots,n^\ast$, we have the following differential equation
\be
\label{eq:113}
\prod_{i=1}^{n^\ast}\left(\frac{d^2}{dt^2}+\omega_i^2\right)\tau^\ast=0.
\ee
This equation can be parameterized with respect to parameters $\theta_i$, $i=1,\dots,n^\ast$, which are functions of $\omega_i^2$, $i=1,\dots,n^\ast$. For instance, in the case that $n^\ast=3$, we have that
\begin{align}
\theta_1=&\omega_1^2\omega_2^2\omega_3^2\nn\\
\theta_2=&\omega_1^2\omega_2^2+\omega_1^2\omega_3^2+\omega_2^2\omega_3^2\nn\\
\theta_3=&\omega_1^2+\omega_2^2+\omega_3^2
\end{align}
which can be considered to be an extension of the standard binomial expansion, namely the expansion of the product of nonidentical binomial polynomials. Then, equation (\ref{eq:113}) can be expanded as
\be
\theta_1\tau^\ast+\theta_2\frac{d^2\tau^\ast}{dt^2}+\theta_3\frac{d^4\tau^\ast}{dt^4}+\frac{d^6\tau^\ast}{dt^6}=0.
\ee
In the general case, the expansion of (\ref{eq:113}) yields
\be
\label{eq:a2}
\theta_1\tau^\ast+\theta_2\frac{d^2\tau^\ast}{dt^2}+\dots+\theta_{n^\ast}\frac{d^{2n^\ast-2}\tau^\ast}{dt^{2n^\ast-2}}+\frac{d^{2n^\ast}\tau^\ast}{dt^{2n^\ast}}=0,
\ee
which involves the same number of unknown parameters as that of the frequencies and is the basis for designing the adaptive differential-cascaded structures to attenuate a disturbance with $n^\ast$ unknown frequencies. We define a stacked reference dynamic system as
\begin{align}
\label{eq:117}
\ddot \chi_1=&\ddot q_d^\ast-\alpha_1^{\ast\ast}\Delta \dot q^\ast-\alpha_0^{\ast\ast}\Delta q^\ast\\
\label{eq:118}
\frac{d^{2n^\ast}\chi_2}{dt^{2n^\ast}}=&\frac{d^{2n^\ast}\chi_1}{d t^{2n^\ast}}-\alpha_{2n^\ast-1}^\ast\frac{d^{2n^\ast-1}\psi_1}{dt^{2n^\ast-1}} \nn\\
&-\alpha_{2n^\ast-2}^\ast\frac{d^{2n^\ast-2}\psi_1}{dt^{2n^\ast-2}} -\dots-\alpha_0^\ast\psi_1\nn\\
&+H^{\ast-1}(p)\{[H^\ast(p)(q-\chi_2)]\hat\theta_1\}\nn\\
&+H^{\ast-1}(p)\{[H^\ast(p)(\ddot q-\ddot \chi_2)]\hat\theta_2\}+\cdots
\nn\\
&+H^{\ast-1}(p)\left\{\left[H^\ast(p)\frac{d^{2n^\ast-2}(q-\chi_2)}{dt^{2n^\ast-2}}\right]\hat\theta_{n^\ast}\right\}
\\
\label{eq:119}
\frac{d^\ell z}{dt^\ell}=&\frac{d^{\ell+1} \chi_2}{dt^{\ell+1}}-\alpha_\ell \frac{d^\ell \psi_2}{dt^\ell}-\alpha_{\ell-1} \frac{d^{\ell-1} \psi_2}{dt^{\ell-1}}\nn\\
&-\dots-\alpha_0\psi_2+\Lambda \frac{d^{\ell-1}}{dt^{\ell-1}}(\dot q-z)
\end{align}
with $H^\ast(p)$ being defined as
\be
\label{eq:a3}
H^\ast(p)=\frac{p^2+\kappa_1^\ast p+\kappa_0^\ast}{p^{2n^\ast}+\alpha_{2n^\ast-1}^\ast p^{2n^\ast-1}+\dots+\alpha_0^\ast},
\ee
and $\hat\theta_i$ the estimate of $\theta_i$, $i=1,\dots,n^\ast$.
The adaptation laws for $\hat \theta_i$, $i=1,\dots,n^\ast$ can be specified in a similar way as above, and in particular
\be
\label{eq:120}
\dot{\hat\theta}_i=-\gamma_i W_i^T(\dot \psi_1+\kappa^\ast \psi_1)
\ee
where $W_i=H^\ast(p)\frac{d^{2i-2}\psi_2}{dt^{2i-2}}$ and $\gamma_i$ is a positive design constant. The vector $h$ and the matrix $W^\ast$ are redefined as
\begin{align}
\label{eq:a4}
h=&\big\{G_1^\ast(p)H^{\ast-1}(p)[\Sigma_{i=1}^{n^\ast}\hat\theta_ip^{2i-2}H^\ast(p)G_2^\ast(p)\nn\\
&\times\Lambda (pI_n+K)^{-1}Y+p^{2n^\ast}H^\ast(p)G_2^\ast(p)\nn\\
&\times\Lambda(pI_n+ K)^{-1}Y]\big\}\hat\vartheta-G_1^\ast(p)H^{\ast-1}(p)\nn\\
&\times[\Sigma_{i=1}^{n^\ast}\hat\theta_ip^{2i-2}H^\ast(p)G_2^\ast(p)\Lambda (pI_n+K)^{-1}(Y\hat\vartheta)\nn\\
&+p^{2n^\ast}H^\ast(p)G_2^\ast(p)\Lambda(pI_n+ K)^{-1}(Y\hat\vartheta)]\\
\label{eq:a5}
W^\ast=&G_1^\ast(p)H^{\ast-1}(p)[\Sigma_{i=1}^{n^\ast}\hat\theta_ip^{2i-2}H^\ast(p)G_2^\ast(p) \nn\\
&\times\Lambda(pI_n+K)^{-1}Y+p^{2n^\ast}H^\ast(p)G_2^\ast(p)\nn\\
&\times\Lambda (pI_n+K)^{-1}Y]
\end{align}
with $G_1^\ast(p)=\frac{p^2}{p^{2n^\ast}+\alpha_{2n^\ast-1}^\ast p^{2n^\ast-1}+\dots+\alpha_0^\ast}$.

\emph{Theorem 9:} Suppose that the disturbance torque $\tau^\ast$ and its derivative of arbitrary order are bounded and that the remainder $
R_t^\ast=G_1^\ast(p)H^{\ast-1}(p)[\Sigma_{i=1}^{n^\ast}\hat\theta_ip^{2i-2}H^\ast(p)G_2^{\ast\ast}(p)R_t+p^{2n^\ast}H^\ast(p)G_2^{\ast\ast}(p)R_t]\to 0
$ as $\ell\to\infty$ and $R_{\psi_1}=H^\ast(p)(\theta_1\psi_2+\theta_2 \ddot \psi_2+\dots+\frac{d^{2n^\ast}}{dt^{2n^\ast}} \psi_2)$ exponentially converges to zero. Then, the forwardstepping adaptive controller given as (\ref{eq:117}), (\ref{eq:118}), (\ref{eq:119}), (\ref{eq:35}), (\ref{eq:36}), (\ref{eq:105}), and (\ref{eq:120}) with $H^\ast(p)$, $h$, and $W^\ast$ being defined as (\ref{eq:a3}), (\ref{eq:a4}), and (\ref{eq:a5}), respectively ensures the convergence of the joint tracking errors as $\ell\to\infty$.

The proof of Theorem 9 can be directly completed with similar procedures as previously conducted.

\emph{Remark 8:} The implementation of the reference dynamics (\ref{eq:92}) and (\ref{eq:118}) can be conducted using degree reduction without involving the measurement of high-order quantities such as $\ddot q$. As compared with the case not involving adaptation to the unknown frequencies of the disturbances, the complexity of the implementation here would be prominently increased since it involves the numerical integration of many dynamical systems. The benefit with adaptive differential-cascaded interconnection is asymptotic compensation of the periodical part of the disturbance using a finite-order differential-cascaded structure while the employment of the previous differential-cascaded interconnection needs to involve an infinite differential series with potentially high gains (see Remark 4).

\emph{Remark 9:} The differential equations (\ref{eq:113}) and (\ref{eq:a2}) describe the properties of the disturbance $\tau^\ast$ with $n^\ast$ frequencies using minimal unknown parameters, which coincide with the result in \cite{Huang2001_TAC} in the context of the internal model methodology. The internal model approach in the literature (e.g., \cite{Wu2019_ICCA}) typically relies on the parametrization of the control input or part of the control input as in \cite{Huang2001_TAC} and the solutions to linear/nonlinear regulator equations. Such constraints render the adaptation of the internal model approach with respect to the uncertain dynamics and unknown disturbances intertwined, yielding the consequence that the adaptation tends to be unreliable and even fragile, similar to the standard adaptive control (e.g., \cite{Ioannou1996_Book,Slotine1991_Book}) as confronting sustaining disturbances. Our result provides a reliable solution to this long-standing problem via exploiting differential-cascaded structures and properties of the sum of differential sequences with a remainder. This reliability is ensured by suitably separating the adaptation to the uncertainty of the system dynamics and that to the disturbance using differential-cascaded structures. The adaptation to the unknown frequencies in the desired trajectories can be implemented using an adaptive differential-cascaded structure in a similar way, based upon the differential equation (\ref{eq:a2}) with $\tau^\ast$ being replaced with the desired position $q_d$, and degree-reduction implementation of the reference dynamics similar to that in Sec. 3.

\emph{Remark 10:} Our result demonstrates a class of differential and integral operations concerning an object that is referred to as a differential function (which involves a function and its derivatives; for instance, Hurwitz polynomials concerning the tracking error $\Delta q$ are differential functions with respect to $\Delta q$), differing from the standard calculus that concerns functions; in addition, the operations are conducted along controlled dynamics around the input. We refer to this class of operations as cascaded calculus (of differential functions along a differential equation). The order of a differential function can be defined in a similar way as that of the standard differential equation. For example, $\Delta \dot q+\alpha\Delta q$ is a differential function (with respect to the function $\Delta q$), and its order is one. The standard function can be considered as a differential function with order zero. A Hurwitz polynomial with respect to $\Delta q$ with degree $m^\ast$ yields a differential function with order $m^\ast$ or the sum of a differential sequence.

\section{Conclusion}

In this paper, we have established a differential-cascaded framework for control of robot manipulators subjected to time-varying disturbances and time-varying desired trajectories. Adaptation in the presence of sustaining disturbances is a long-standing challenging problem for several decades, and to the best of our knowledge, the reliable solution has not yet been systematically developed. One contribution of our result is to provide a reliable solution to this problem, which relies on the exploitation of high-order differential-cascaded structures for low-order robot manipulators involving infinite-order reference dynamics or infinite differential series. We also accommodate the trajectory tracking of robot manipulators with partial knowledge of the desired trajectory (e.g., only the desired position is available) via exploiting differential-cascaded structures. As indicated in \cite{Wang2020_arXiv}, the differential-cascaded approach is a constructive one with the interconnection or cascade component involving the derivative or high-order derivatives of the states of the system, differing from the conventional cascaded approach (which can also be considered to be a differential-cascaded approach with degree zero). Our study here demonstrates the effective application of the differential-cascaded paradigm to solving long-standing problems of sustaining interest in systems and control via the standard example of nonlinear robot manipulators. We also witness the application of a systematic tool for handling nonlinearity, uncertainty, and robustness issue in control systems, namely the nonlinear control problem is transformed to the one concerning a differential equation involving the sum of a high-order differential sequence with a remainder.

In our present study, we have observed a class of mathematical operations concerning differential functions, the application of which has also been witnessed in \cite{Wang2020_AUT,Wang2020b_AUT,Wang2019_ACC,Wang2019_CDC,Wang2020_arXiv}, and the most prominent example may be the one involving the introduction of stacked reference dynamics. This class of differential and integral operations is referred to as cascaded calculus (of differential functions along a differential equation), in contrast to the standard calculus that involves functions; in the context of control, we can refer to this class of operations as cascaded calculus along controlled dynamics around the input. As is well known, one significant mathematical object due to the introduction of calculus is a differential equation, which plays an important role in describing the evolution of many physical systems with or without an external input. The application of cascaded calculus or cascaded calculus of differential functions along a differential equation yields a structure governing dynamical systems that is referred to as a differential-cascaded structure \cite{Wang2020_arXiv}; constructing differential-cascaded structures for dynamical systems demonstrates its instrumental role in providing solutions to many important control problems.

%\section*{Acknowledgment}

\bibliographystyle{plain}        % Include this if you use bibtex
\bibliography{..//Reference_list_Wang}           % and a bib file to produce the

\begin{thebibliography}{10}

\bibitem{Astolfi2003_TAC}
A.~Astolfi and R.~Ortega.
\newblock Immersion and invariance: {A} new tool for stabilization and adaptive
  control of nonlinear systems.
\newblock {\em IEEE Transactions on Automatic Control}, 48(4):590--606, Apr.
  2003.

\bibitem{Berghuis1993_SCL}
H.~Berghuis and H.~Nijmeijer.
\newblock Global regulation of robots using only position measurements.
\newblock {\em Systems \& Control Letters}, 21(4):289--293, Oct. 1993.

\bibitem{Byrnes2000_IJRNC}
C.~I. Byrnes and A.~Isidori.
\newblock Output regulation for nonlinear systems: An overview.
\newblock {\em International Journal of Robust and Nonlinear Control},
  10(5):323--337, Apr. 2000.

\bibitem{Cao2012_TAC}
Y.~Cao and W.~Ren.
\newblock Distributed coordinated tracking with reduced interaction via a
  variable structure approach.
\newblock {\em IEEE Transactions on Automatic Control}, 57(1):33--48, Jan.
  2012.

\bibitem{Cheah1999_ICRA}
C.~C. Cheah, S.~Kawamura, S.~Arimoto, and K.~Lee.
\newblock {PID} control of robotic manipulator with uncertain {J}acobian
  matrix.
\newblock In {\em Proceedings of the IEEE International Conference on Robotics
  and Automation}, Detroit, Michigan, USA, 1999.

\bibitem{Cheah2006_IJRR}
C.~C. Cheah, C.~Liu, and J.-J.~E. Slotine.
\newblock Adaptive tracking control for robots with unknown kinematic and
  dynamic properties.
\newblock {\em The International Journal of Robotics Research}, 25(3):283--296,
  Mar. 2006.

\bibitem{Chen2007_CDC}
Z.~Chen and J.~Huang.
\newblock An adaptive regulation problem and its application to spacecraft
  systems.
\newblock In {\em Proceedings of the IEEE Conference on Decision and Control},
  pages 4631--4636, New Orleans, LA, USA, 2007.

\bibitem{Craig1987_IJRR}
J.~J. Craig, P.~Hsu, and S.~S. Sastry.
\newblock Adaptive control of mechanical manipulators.
\newblock {\em The International Journal of Robotics Research}, 6(2):16--28,
  Jun. 1987.

\bibitem{Desoer1975_Book}
C.~A. Desoer and M.~Vidyasagar.
\newblock {\em Feedback Systems: Input-Output Properties}.
\newblock Academic Press, New York, 1975.

\bibitem{Huang2001_TAC}
J.~Huang.
\newblock Remarks on the robust output regulation problem for nonlinear
  systems.
\newblock {\em IEEE Transactions on Automatic Control}, 46(12):2028--2031, Dec.
  2001.

\bibitem{Ioannou1996_Book}
P.~A. Ioannou and J.~Sun.
\newblock {\em Robust Adaptive Control}.
\newblock Prentice-Hall, Upper Saddle River, NJ, 1996.

\bibitem{Isidori1990_TAC}
A.~Isidori and C.~I. Byrnes.
\newblock Output regulation of nonlinear systems.
\newblock {\em IEEE Transactions on Automatic Control}, 35(2):131--140, Feb.
  1990.

\bibitem{Jayawardhana2008_AUT}
B.~Jayawardhana and G.~Weiss.
\newblock Tracking and disturbance rejection for fully actuated mechanical
  systems.
\newblock {\em Automatica}, 44(11):2863--2868, Nov. 2008.

\bibitem{Kelly1993_IFAC}
R.~Kelly.
\newblock A simple set-point robot controller by using only position
  measurements.
\newblock In {\em IFAC World Congress}, pages 527--530, Sydney, Australia,
  1993.

\bibitem{Kokotovic2001_AUT}
P.~V. Kokotovi{\'c} and M.~Arcak.
\newblock Constructive nonlinear control: A historical perspective.
\newblock {\em Automatica}, 37(5):637--662, May 2001.

\bibitem{Krstic1995_Book}
M.~Krsti{\'c}, I.~Kanellakopoulos, and P.~V. Kokotovi{\'c}.
\newblock {\em Nonlinear and Adaptive Control Design}.
\newblock Wiley, New York, 1995.

\bibitem{Krstic1995_TAC}
M.~Krsti{\'c} and P.~V. Kokotovi{\'c}.
\newblock Adaptive nonlinear design with controller-identifier separation and
  swapping.
\newblock {\em IEEE Transactions on Automatic Control}, 40(3):426--440, Mar.
  1995.

\bibitem{Liu2006_TRO}
Y.-H. Liu, H.~Wang, C.~Wang, and K.~K. Lam.
\newblock Uncalibrated visual servoing of robots using a depth-independent
  interaction matrix.
\newblock {\em IEEE Transactions on Robotics}, 22(4):804--817, Aug. 2006.

\bibitem{Lu2019_AUT}
M.~Lu, L.~Liu, and G.~Feng.
\newblock Adaptive tracking control of uncertain {E}uler--{L}agrange systems
  subject to external disturbances.
\newblock {\em Automatica}, 104:207--219, Jun. 2019.

\bibitem{Middleton1988_SCL}
R.~H. Middleton and G.~C. Goodwin.
\newblock Adaptive computed torque control for rigid link manipulators.
\newblock {\em Systems \& Control Letters}, 10(1):9--16, Jan. 1988.

\bibitem{Ortega1989_AUT}
R.~Ortega and M.~W. Spong.
\newblock Adaptive motion control of rigid robots: {A} tutorial.
\newblock {\em Automatica}, 25(6):877--888, Nov. 1989.

\bibitem{Patre2010_AUT}
P.~M. Patre, W.~MacKunis, M.~Johnson, and W.~E. Dixon.
\newblock Composite adaptive control for {E}uler--{L}agrange systems with
  additive disturbances.
\newblock {\em Automatica}, 46(1):140--147, Jan. 2010.

\bibitem{Sepulchre1997_AUT}
R.~Sepulchre, M.~Jankovi{\'c}, and P.~V. Kokotovi{\'c}.
\newblock Integrator forwarding: A new recursive nonlinear robust design.
\newblock {\em Automatica}, 33(5):979--984, May 1997.

\bibitem{Slotine1987_IJRR}
J.-J.~E. Slotine and W.~Li.
\newblock On the adaptive control of robot manipulators.
\newblock {\em The International Journal of Robotics Research}, 6(3):49--59,
  Sep. 1987.

\bibitem{Slotine1989_AUT}
J.-J.~E. Slotine and W.~Li.
\newblock Composite adaptive control of robot manipulators.
\newblock {\em Automatica}, 25(4):509--519, Jul. 1989.

\bibitem{Slotine1991_Book}
J.-J.~E. Slotine and W.~Li.
\newblock {\em Applied Nonlinear Control}.
\newblock Prentice-Hall, Englewood Cliffs, NJ, 1991.

\bibitem{Spong1989_SCL}
M.~W. Spong.
\newblock Adaptive control of flexible joint manipulators.
\newblock {\em Systems \& Control Letters}, 13(1):15--21, Jul. 1989.

\bibitem{Spong2006_Book}
M.~W. Spong, S.~Hutchinson, and M.~Vidyasagar.
\newblock {\em Robot Modeling and Control}.
\newblock Wiley, Hoboken, NJ, 2006.

\bibitem{Takegaki1981_ASME}
M.~Takegaki and S.~Arimoto.
\newblock A new feedback method for dynamic control of manipulators.
\newblock {\em Journal of Dynamic Systems, Measurement, and Control},
  103(2):119--125, Jun. 1981.

\bibitem{Teel1992_IFAC}
A.~R. Teel.
\newblock Using saturation to stabilize a class of single-input partially
  linear composite systems.
\newblock In {\em IFAC Nonlinear Control Systems Design}, pages 379--384,
  Bordeaux, France, 1992.

\bibitem{Wang2015_AUT}
H.~Wang.
\newblock Adaptive visual tracking for robotic systems without image-space
  velocity measurement.
\newblock {\em Automatica}, 55:294--301, May 2015.

\bibitem{Wang2017_TAC}
H.~Wang.
\newblock Adaptive control of robot manipulators with uncertain kinematics and
  dynamics.
\newblock {\em IEEE Transactions on Automatic Control}, 62(2):948--954, Feb.
  2017.

\bibitem{Wang2019_CDC}
H.~Wang.
\newblock Task-space bilateral control of teleoperators with time-varying
  delay.
\newblock In {\em Proceedings of the IEEE Conference on Decision and Control},
  pages 1698--1703, Nice, France, 2019.

\bibitem{Wang2020_AUT}
H.~Wang.
\newblock Differential-cascade framework for consensus of networked
  {L}agrangian systems.
\newblock {\em Automatica}, 112:108620, Feb. 2020.

\bibitem{Wang2020b_AUT}
H.~Wang.
\newblock Towards manipulability of interactive {L}agrangian systems.
\newblock {\em Automatica}, 119:108913, Sep. 2020.

\bibitem{Wang2021_arXiv}
H.~Wang.
\newblock A differential-cascaded approach for adaptive control of robot
  manipulators.
\newblock {\em arXiv preprint arXiv: 2104.03853}, Apr. 2021.

\bibitem{Wang2019_ACC}
H.~Wang, W.~Ren, and C.~C. Cheah.
\newblock Forwardstepping: {A} new approach for control of dynamical systems.
\newblock In {\em Proceedings of the American Control Conference}, pages
  1208--1215, Philadelphia, PA, USA, 2019.

\bibitem{Wang2020_arXiv}
H.~Wang, W.~Ren, and C.~C. Cheah.
\newblock A differential-cascaded paradigm for control of nonlinear systems.
\newblock {\em arXiv preprint arXiv:2012.14251}, Dec. 2020.

\bibitem{Wang2020_TAC}
H.~Wang, W.~Ren, C.~C. Cheah, Y.~Xie, and S.~Lyu.
\newblock Dynamic modularity approach to adaptive control of robotic systems
  with closed architecture.
\newblock {\em IEEE Transactions on Automatic Control}, 65(6):2760--2767, Jun.
  2020.

\bibitem{Wu2019_ICCA}
H.~Wu and D.~Xu.
\newblock Inverse optimality and adaptive asymptotic tracking control of
  uncertain {E}uler-{L}agrange systems.
\newblock In {\em IEEE International Conference on Control and Automation},
  pages 242--247, Edinburgh, U.K., 2019.

\bibitem{Wu2020_CCC}
H.~Wu, D.~Xu, and B.~Jayawardhana.
\newblock Output regulation of {E}uler-{L}agrange systems based on error and
  velocity feedback.
\newblock In {\em Proceedings of the Chinese Control Conference}, pages
  604--609, Shenyang, China, 2020.

\bibitem{Xian2004_TAC}
B.~Xian, D.~M. Dawson, M.~S. de~Querioz, and J.~Chen.
\newblock A continuous asymptotic tracking control strategy for uncertain
  nonlinear systems.
\newblock {\em IEEE Transactions on Automatic Control}, 49(7):1206--1211, Jul.
  2004.

\end{thebibliography}
                                 % bibliography (preferred). The
                                 % correct style is generated by
                                 % Elsevier at the time of printing.
\end{document}